\documentclass[12pt]{article}

\usepackage{scicite}

\usepackage{times}
\usepackage{longtable}
\usepackage{graphicx}

\newcommand{\addition}[1]{#1}
\newcommand{\deletion}[1]{}

\newenvironment{sciabstract}{%
\begin{quote} \bf}
{\end{quote}}

\newcounter{lastnote}

\title{Kepler-36: A Pair of Planets with Neighboring Orbits and Dissimilar Densities}

\author{
Joshua A. Carter$^1$+$^\ast$,
Eric Agol$^2$+$^\ast$,
William J. Chaplin$^{3}$,
\\
Sarbani Basu$^{4}$,
Timothy R. Bedding$^{5}$,
Lars A. Buchhave$^{6}$,
\\
J\o rgen Christensen-Dalsgaard$^{7}$,
Katherine M. Deck$^{8}$,
Yvonne Elsworth$^{3}$,
\\
Daniel C. Fabrycky$^{9}$,
Eric B. Ford$^{10}$,
Jonathan J. Fortney$^{11}$,
\\
Steven J. Hale$^{3}$,
Rasmus Handberg$^{7}$,
Saskia Hekker$^{12}$,
\\
Matthew J. Holman$^{13}$,
Daniel Huber$^{14}$,
Christopher Karoff$^{7}$,
\\
Steven D. Kawaler$^{15}$,
Hans Kjeldsen$^{7}$,
Jack J. Lissauer$^{14}$,
\\
Eric D. Lopez$^{11}$,
Mikkel N. Lund$^{7}$,
Mia Lundkvist$^{7}$,
\\
Travis S. Metcalfe$^{16}$,
Andrea Miglio$^{3}$,
Leslie A. Rogers$^{8}$,
\\
Dennis Stello$^{5}$,
William J. Borucki$^{14}$,
Steve Bryson$^{14}$,
\\
Jessie L. Christiansen$^{17}$,
William D. Cochran$^{18}$,
John C. Geary$^{13}$,
\\
Ronald L. Gilliland$^{19}$,
Michael R. Haas$^{14}$,
Jennifer Hall$^{20}$,
\\
Andrew W. Howard$^{21}$,
Jon M. Jenkins$^{17}$,
Todd Klaus$^{20}$,
\\
David G. Koch$^{14}$,
David W. Latham$^{13}$,
Phillip J. MacQueen$^{18}$,
\\
Dimitar Sasselov$^{13}$,
Jason H. Steffen$^{22}$,
Joseph D. Twicken$^{17}$,
\\
Joshua N. Winn$^{8}$
\and
\normalsize{$^{1}$\parbox{6.5in}{Hubble Fellow, Harvard-Smithsonian Center for Astrophysics, 60 Garden Street, Cambridge, MA 02138, USA}}\and
\normalsize{$^{2}$\parbox{6.5in}{Department of Astronomy, Box 351580, University of Washington, Seattle, WA 98195, USA}}\and
\normalsize{$^{3}$\parbox{6.5in}{School of Physics and Astronomy, University of Birmingham, Edgbaston, B15 2TT, UK}}\and
\normalsize{$^{4}$\parbox{6.5in}{Department and Astronomy, Yale University, New Haven, CT, 06520, USA}}\and
\normalsize{$^{5}$\parbox{6.5in}{Sydney Institute for Astronomy, School of Physics, University of Sydney, Sydney, Australia}}\and
\normalsize{$^{6}$\parbox{6.5in}{Niels Bohr Institute, Copenhagen University, DK-2100 Copenhagen, Denmark\\Centre for Star and Planet Formation, Natural History Museum of Denmark, University of Copenhagen, DK-1350 Copenhagen, Denmark}}\and
\normalsize{$^{7}$\parbox{6.5in}{Stellar Astrophysics Centre (SAC), Department of Physics and Astronomy, Aarhus University, Ny Munkegade 120, DK-8000 Aarhus C, Denmark}}\and
\normalsize{$^{8}$\parbox{6.5in}{Massachusetts Institute of Technology, Physics Department and Kavli Institute for Astrophysics and Space Research, 77 Massachusetts Avenue, Cambridge, MA 02139, USA}}\and
\normalsize{$^{9}$\parbox{6.5in}{Hubble Fellow, Department of Astronomy and Astrophysics, University of California, Santa Cruz, CA 95064}}\and
\normalsize{$^{10}$\parbox{6.5in}{Department of Astronomy, University of Florida, Gainesville, FL 32611-2055, USA}}\and
\normalsize{$^{11}$\parbox{6.5in}{Department of Astronomy and Astrophysics, University of California, Santa Cruz, CA 95064}}\and
\normalsize{$^{12}$\parbox{6.5in}{Astronomical Institute `Anton Pannekoek', University of Amsterdam, The Netherlands\\School of Physics and Astronomy, University of Birmingham, Edgbaston, B15 2TT, UK}}\and
\normalsize{$^{13}$\parbox{6.5in}{Harvard-Smithsonian Center for Astrophysics, 60 Garden Street, Cambridge, MA 02138, USA}}\and
\normalsize{$^{14}$\parbox{6.5in}{NASA Ames Research Center, Moffett Field, CA 94035, USA}}\and
\normalsize{$^{15}$\parbox{6.5in}{Department of Physics and Astronomy, Iowa State University, Ames, IA, 50011, USA}}\and
\normalsize{$^{16}$\parbox{6.5in}{White Dwarf Research Corporation, Boulder, CO, 80301, USA}}\and
\normalsize{$^{17}$\parbox{6.5in}{SETI Institute/NASA Ames Research Center, Moffett Field, CA 94035}}\and
\normalsize{$^{18}$\parbox{6.5in}{McDonald Observatory, University of Texas at Austin, Austin, TX, 78712, USA}}\and
\normalsize{$^{19}$\parbox{6.5in}{Center for Exoplanets and Habitable Worlds, The Pennsylvania State University, University Park, PA, 16802, USA}}\and
\normalsize{$^{20}$\parbox{6.5in}{Orbital Science Corporation/NASA Ames Research Center, Moffett Field, CA 94035}}\and
\normalsize{$^{21}$\parbox{6.5in}{Department of Astronomy, University of California, Berkeley, CA, 94720, USA}}\and
\normalsize{$^{22}$\parbox{6.5in}{Fermilab Center for Particle Astrophysics, P.O. Box 500, Batavia IL 60510, USA}}\and
\normalsize{+These authors contributed equally to this work.} \\
\normalsize{$^\ast$To whom correspondence should be addressed. E-mail:}  \\ 
\normalsize{jacarter@cfa.harvard.edu, agol@astro.washington.edu}
}

\date{}

\begin{document} 


\baselineskip24pt


\maketitle 


\begin{sciabstract}

  In the Solar system the planets' compositions vary with orbital distance, with rocky planets in close orbits and lower-density gas giants in wider orbits. 
  The detection of close-in giant planets around other stars was the first clue that this pattern is not universal, and that planets' orbits can change substantially after their formation. Here we report another violation of the orbit-composition pattern: two planets orbiting the same star with orbital distances differing by only 10\%, and densities differing by a factor of 8. One planet is likely a rocky `super-Earth', whereas the other is more akin to Neptune. 
  These planets are twenty times more closely spaced---and have a larger density contrast---than any adjacent pair of planets in the Solar system.

\end{sciabstract}

The detection of the first `hot Jupiter' around a Sun-like star \cite{Mayor3781995} was surprising to researchers who based their expectations on the properties of the Solar system. Theorists soon developed models of planetary `migration' to explain how the planets' orbits can shrink \cite{Lin3801996}. This led to a broader recognition that the architecture of planetary systems can change substantially after their formation. The planetary system reported in this paper is another example of an `extreme' planetary system that will serve as a stimulus to theories of planet migration and orbital rearrangement. The system features two planets in neighboring orbits with substantially different compositions.

This system was discovered from the miniature eclipses or ``transits'' that cause the host star to appear fainter when the planets pass in front of the star.  The target star (Kepler Object of Interest 277; hereafter, Kepler-36; also KIC 11401755, 2MASS 19250004+4913545) is one of approximately 150,000 stars that is subject to nearly continuous photometric surveillance by the Kepler spacecraft \cite{Borucki3272010,Koch7132010,Caldwell7132010}.  The Kepler data revealed the transits of two planets.  The loss of light during transits by planet b is only 17\% as large as that from planet c (Fig 1). Both transit sequences show substantial timing variations (Fig 2). Indeed the variations are large enough that the smaller planet was initially overlooked by the Kepler data reduction pipeline \cite{Jenkins77402010,Jenkins7132010}, which makes the usual assumption of strictly periodic orbits. It was subsequently identified with a search algorithm allowing for variations between successive transits up to a specified fraction of the mean period \cite{Kel'Manov522004}.

The candidates have average orbital periods of $P_b=13.84$ d and $P_c=16.23$ d, which have a ratio close to $6/7$. The transit timing variations are approximately linear in time, with abrupt slope changes after planetary conjunctions ($\approx 6\times P_b \approx 7\times P_c \approx$ every $97$~d). The variations are anti-correlated; when one body is early, the other is late, and vice versa. This is diagnostic of gravitational interactions \cite{Ford2012,Steffen2012,Fabrycky2012}, leaving no doubt that the two transiting bodies orbit the same star (as opposed to other scenarios in which two independent eclipsing systems are spatially unresolved by the telescope).

Even without detailed modeling it is possible to show that the transiting bodies must be planets, by imposing the requirement of orbital stability. For long-term stability against collisions or ejections, the sum of the orbiting bodies' masses must be less than $10^{-4}$ times the mass of the star \cite{Marchal261982,Gladman1061993,Barnes6472006}. This criterion leads to an upper bound of 40 Earth masses, hence planetary masses for both bodies. 

Precise knowledge of the masses and radii of the planets requires precise knowledge of those same properties for the star. This information is typically gleaned from spectroscopic observations. In this case there is an additional source of information: the Kepler data reveal solar-like oscillations of the host star, due to excitation of sound waves by turbulence in the stellar atmosphere. The frequencies of the oscillations indicate that the star has a density $(25\pm 2)\%$ that of the Sun \cite{Chaplin3322011, SOM}.  Analysis of high-resolution spectra of this star, subject to this density constraint, yields precise values for the stellar effective temperature and metallicity.  The star is slightly hotter and less metal-rich than the Sun.  This information combined with additional asteroseismic constraints \cite{SOM} gives precise measures of the stellar mass and radius (Table 1). Based on these parameters \cite{SOM}, Kepler-36 is a 'subgiant' star, 2-3 billion years older than the Sun.

We used the seismic results as constraints on the parameters of a photometric-dynamical model fitted to the Kepler data \cite{SOM}. This model is based on the premise of a star and two planets interacting via Newtonian gravity with an appropriate loss of light whenever a planet is projected in front of the star \cite{Carter3312011}. Given initial positions and masses, the positions of the three bodies at the observed times were computed numerically, and the loss of light was computed assuming a linear limb-darkening law \cite{Mandel5802002}.  Only data within half a day of a transit were fitted, after correcting for extraneous trends by removing a linear function of time \cite{SOM}.

We adjusted the parameters of the model to fit to the data. The optimal solution agrees well with the data, in particular with the transit-timing variations of both planets (Fig 1 and Fig 2). We calculated the credible intervals for the model parameters \cite{SOM} with a Differential Evolution Markov Chain Monte Carlo algorithm \cite{terBraak} (Table 1).

The masses and radii have uncertainties less than $8\%$ and $3\%$ respectively. This unusually good precision owes to the combination of the constraint on the planetary mass ratio ($0.55\pm0.01$) and planets-to-star mass ratio [($3.51 \pm 0.20) \times 10^{-5}$] obtained from the detection of transit-timing variations, coupled with the usual geometric transit parameters and the asteroseismic constraints \cite{SOM}.

Knowledge of the planetary masses and radii enables the study of their possible compositions \cite{SOM}.  Planet b's dimensions are consistent with a rocky Earth-like composition, with approximately $30\%$ of its mass in iron.  Any volatile constituents --- those having relatively low boiling points such as hydrogen, helium or water --- must be a small fraction of b's volume.  As a limiting case, if the planet is assumed to have a maximally iron-rich core \cite{Marcus7002009}, water may not constitute more than $15\%$ of the total mass, and any H/He atmosphere contributes less than $1\%$ of the total mass. In contrast, planet c must be volatile-rich; it is less dense than water, implying a substantial H/He atmosphere. Taking the interior to be free of water, with an Earth-like composition, the H/He atmosphere would contain 9\% of the total mass. Even if the planet were half water by mass, the H/He atmosphere would still be $>1$\% of the total mass (Fig 3).

As for the orbits, the photometric-dynamical model shows \cite{SOM} they are nearly circular (eccentricities $<0.04$) and coplanar (mutual inclination $<2.5$~degrees). The evidence for coplanarity comes from the lack of significant variation in the transit durations \cite{SOM,Miralda-Escud&eacute;5642002}. The orbits are also closely spaced: at conjunctions the planets approach one another within 0.013 AU. The angular size of c as viewed from b would be 2.5 times larger than the full Moon viewed from Earth. We investigated whether these close encounters are consistent with dynamical stability by scrutinizing a random sample of allowed model parameters \cite{SOM}. Direct numerical integration \cite{SOM} showed that more than 91\% of this sample avoided disruptive encounters over $\sim$0.7 million years. From this surviving population, 100 parameter sets were drawn randomly and numerically integrated for 140 million years; none experienced disruptive encounters.

Of all the multiplanet systems for which densities of the planets have been measured, Kepler-36 has the smallest fractional separation between any pair of adjacent orbits, and this pair also has one of the largest density contrasts (Fig S20). These factors present an interesting problem. One would naturally suppose that the two planets formed at widely separated locations in the protoplanetary disk, with volatile-poor b inside the `snow line' at a few AU, and volatile-rich c outside. It will be interesting to see whether the usual `migration' mechanism that is invoked to alter planetary orbits --- tidal interactions with the gaseous protoplanetary disk --- could draw together two planets from such different regions of the disk. Or whether the compositions and densities of the planets could have changed with time, due for example to the preferential erosion of the smaller planet's atmosphere by stellar irradiation \cite{SOM} (Fig S21). Perhaps a combination of these factors will ultimately explain this puzzling pair of planets.

\paragraph*{\bf Acknowledgements}

NASAÕs Science Mission Directorate provided funding for the Kepler Discovery mission. JAC and DCF acknowledge support by NASA through Hubble Fellowship grants HF-51267.01-A and HF-51272.01-A awarded by the Space Telescope Science Institute, which is operated by the Association of Universities for Research in Astronomy, Inc., for NASA, under contract NAS 5-26555.  EA acknowledges NSF Career grant AST-0645416, and thanks the Center for Astrophysics where this work began. WJC, AM and YE acknowledge the financial support of the UK Science and Technology Facilities Council (STFC). Funding for the Stellar Astrophysics Centre (SAC) is provided by The Danish
National Research Foundation. The research is supported by the ASTERISK
project (ASTERoseismic Investigations with SONG and Kepler) funded by the
European Research Council (Grant agreement no.: 267864). SH acknowledges financial support from the Netherlands Organisation for Scientific Research (NWO).  Computational time on Kraken at the National Institute of Computational Sciences was provided through NSF TeraGrid allocation TG-AST090107. J.N.W. was supported by the NASA Kepler Participating Scientist
program through grant NNX12AC76G.  

Refer to the supplemental material for access information to data utilized in this work.

\clearpage

\begin{table}
\centering
\begin{tabular}{lc}
{\it Star} & \\ \hline
~Mass, $M_\star$ ($M_\odot$) & $1.071\pm0.043$\\
~Radius, $R_\star$ ($R_\odot$) & $1.626\pm0.019$\\
~Mean Density, $\rho_\star$ (g cm$^{-3}$) & $0.3508\pm0.0056$\\
~Stellar Effective Temperature, $T_{\rm eff}$ (K) & $5911 \pm 66$ \\
{\it Planet b} & \\ \hline
~Time of Transit, $T_b$ (BJD) & $2454960.9753_{-      0.0058}^{+      0.0055}$ \\
~Period, $P_b$ (day) & $13.83989_{- 0.00060}^{+ 0.00082}$ \\
~Orbital Semimajor Axis, $a_b$ (AU) & $0.1153\pm0.0015$\\
~Mass, $M_b$ ($M_\oplus$) & $4.45_{-0.27}^{+0.33}$ \\
~Radius, $R_b$ ($R_\oplus$) & $1.486\pm0.035$\\
~Mean Density, $\rho_b$ (g cm$^{-3}$) & $7.46_{-0.59}^{+0.74}$ \\
~Equilibrium Temperature, $T_{\rm eq, b}$ (K) & $978\pm11$\\
{\it Planet c} & \\ \hline
~Time of Transit, $T_c$ (BJD) & $2454955.9132_{-      0.0010}^{+      0.0011}$ \\
~Period, $P_c$ (day) & $16.23855_{- 0.00054}^{+ 0.00038}$ \\
~Orbital Semimajor Axis, $a_c$ (AU) & $0.1283\pm0.0016$\\
~Mass, $M_c$ ($M_\oplus$) & $8.08_{-0.46}^{+0.60}$ \\
~Radius, $R_c$ ($R_\oplus$) & $3.679\pm0.054$\\
~Mean Density, $\rho_c$ (g cm$^{-3}$) & $0.89_{-0.05}^{+0.07}$ \\
~Equilibrium Temperature, $T_{\rm eq, c}$ (K) & $928\pm10$\\
\end{tabular}
\caption{{\bf Characteristics of the Kepler-36 system.} The stellar parameters
are based on the analysis of the optical spectrum and the
asteroseismic oscillations observed in the {\it Kepler} data \cite{SOM}. ÊBecause the orbits are not strictly periodic, the periods and times of transit given
here refer to instantaneous (`osculating') values evaluated at an
arbitrary reference epoch (2,454,950 Barycentric Julian Date, BJD).
The parameter ranges are based on the medians and 68\% confidence
limits of the marginalized posteriors. The equilibrium
temperature was calculated assuming a Bond albedo of $A_B = 0.3$, with
$T_{\rm eq} = T_{\rm eff} \sqrt{R_\star/2 a}
(1-A_B)^{1/4}$.  }
\end{table}

\clearpage

\begin{figure}
\center
\includegraphics{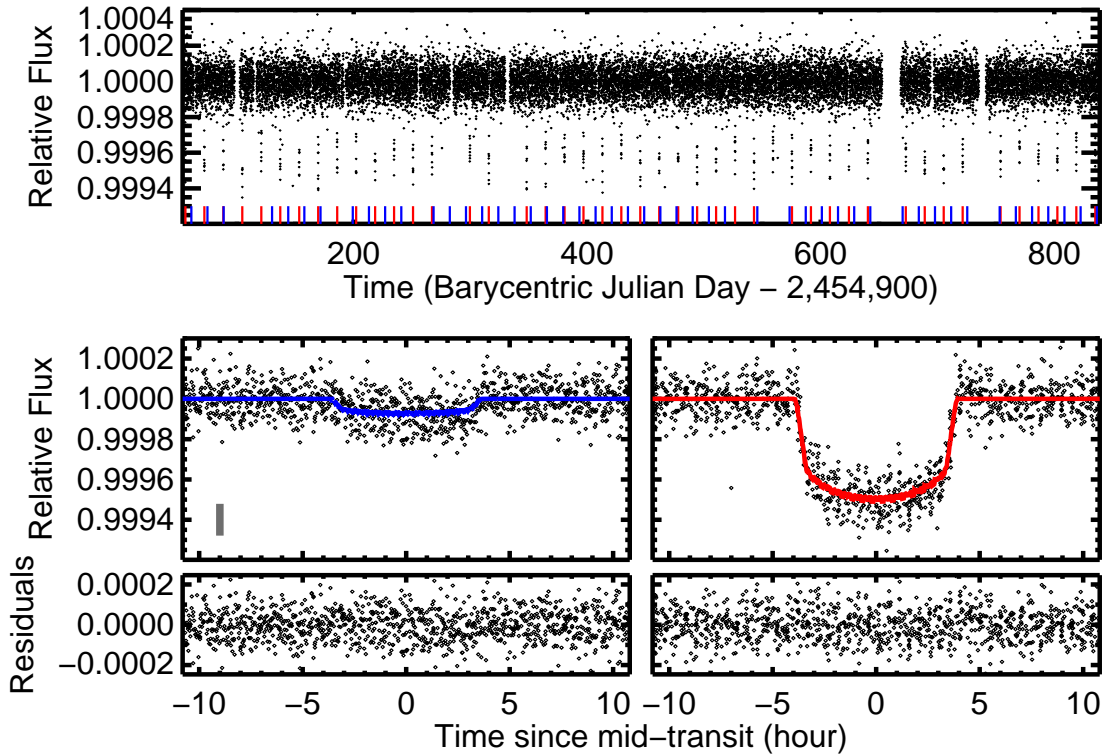}
\caption{{\bf Kepler-36 light curve.} Ê{\bf Top.---}Relative flux over 877
days, after applying a median filter to remove long-term trends caused by
astrophysical and instrumental effects. ÊThe brief 0.05\% dips
represent transits of planet c (also marked with red bars on the time
axis). ÊThe individual transits of planet b (blue bars) are not easily
discerned in the time series. The alternating pattern of transits (blue, red, blue,...)  is flipped every $\approx 97$ days as the planets pass each other in their orbits. {\bf Bottom.---}Composite transit light
curves and best-fitting models, obtained by subtracting from each time stamp the nearest
mid-transit time (calculated from the best-fitting model),  and the data residuals. The gray bar shows the typical scatter of the relative flux.  The transit
of planet c (right panel, model in red) results in a loss of light about 6 times
greater than that due to planet b (left panel, model in blue). } \label{fig02}
\end{figure}

\begin{figure}
\center
\includegraphics{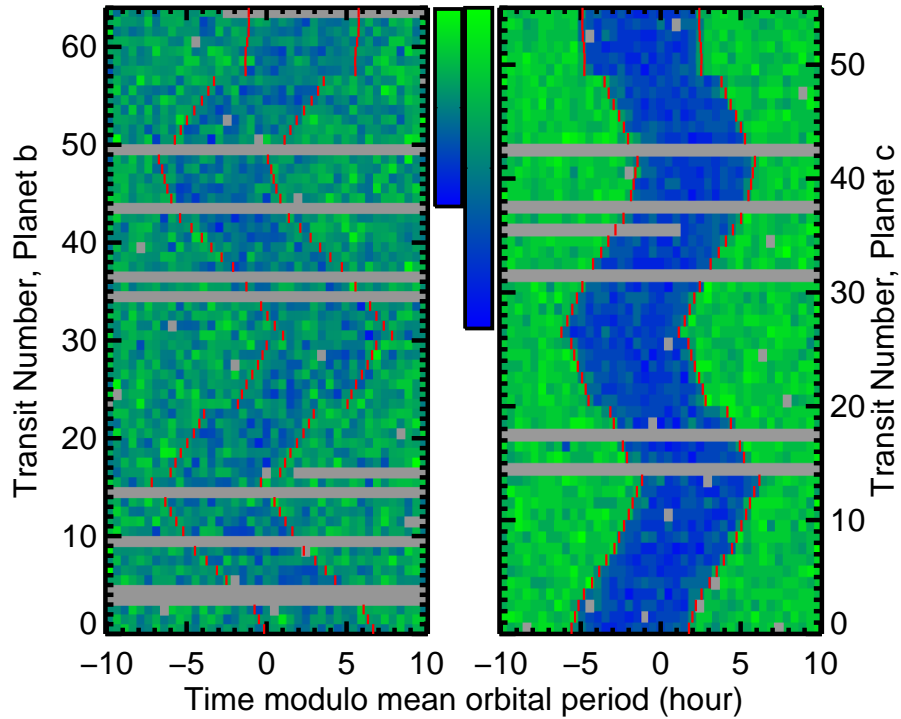}
\caption{{\bf Visualization of transit-timing perturbations.} ÊColored pixels
encode relative flux (increasing from blue to green), with light gray
pixels representing unavailable data. ÊEach row represents an
individual transit light curve, from the earliest observed transit at
the bottom to the most recent transit at the top. ÊThe horizontal axis
is the time modulo the mean orbital period for that planet. ÊStrictly
periodic transits would produce a blue vertical band. ÊThe curved,
riverine bands that are observed are indicative of transit-timing
perturbations. ÊThe red ticks on each row indicate the start and end time of transit, according to the best-fitting dynamical model. 
The range of relative fluxes in each panel spans the minimum and maximum data values in the middle panels of Figure~\ref{fig02}.  } \label{fig03}
\end{figure}

\begin{figure}
\center
\includegraphics{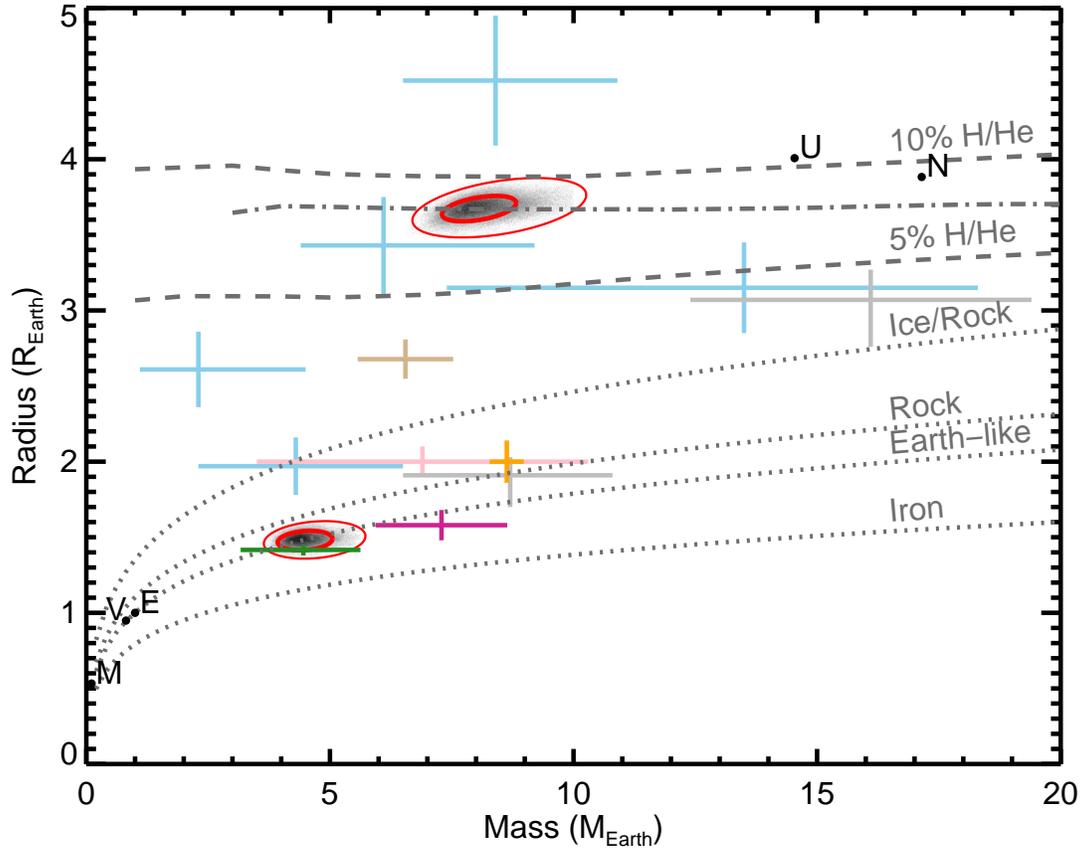}
\caption{{\bf Mass-radius diagram for small planets.} ÊConstraints for
Kepler-36b and c are shown as two-dimensional joint probability
densities and confidence contours (68\% and 95\%). ÊOther exoplanets
are shown for comparison: blue -- the planets in Kepler-11
\cite{Lissauer4702011}, pink -- Kepler-18b \cite{Cochran1972011},
gray -- Kepler-20 b and c \cite{Gautier2012}, brown -- GJ 1214b
\cite{Charbonneau4622009}, violet -- CoRoT-7b \cite{Hatzes7432011},
green -- Kepler-10b \cite{Batalha7292011}, orange -- 55 Cnc e \cite{Winn7372011}. Solar System planets are
plotted using the first letter of their names (excluding Mercury). ÊThe curves represent
theoretical models for planets of a given composition. Dotted curves
are models of terrestrial bodies [those lacking a significant gas
envelope \cite{Fortney6592007}], with ``Ice/Rock'' -- 50\% ice and
rock (silicates) by mass, ``Rock'' -- 100\% rock, ``Earth-like'' --
33\% iron, 67\% rock, ``Iron'' -- 100\% iron. ÊDashed curves are for
planets with Earth-like solid cores surrounded with H/He envelopes
with 5\% or 10\% of the total mass. The dash-dotted curve is for an Earth-like core and a water layer, in equal parts mass, surrounded by H/He envelope with 1.6\% of the total mass.  }

\end{figure}

\clearpage

\baselineskip16pt

\setcounter{figure}{0}
\setcounter{table}{0}
\renewcommand{\thefigure}{S\arabic{figure}}
\renewcommand{\thetable}{S\arabic{table}}

\paragraph*{\Huge Supporting Online Material} 

\paragraph{In this supplement, we 
provide additional details regarding the discovery of Kepler-36 (KOI-277, KIC 11401755, 2MASS 19250004+4913545). This supplement is organized as follows. In \S \ref{sec:star}, we describe the spectroscopic reconnaissance (\S  \ref{sec:star:spec}) and asteroseismology (\S \ref{sec:astero}) of Kepler-36 that constrained the stellar properties, summarized in \S \ref{sec:res}.  In \S \ref{sec:details}, we describe how constraints on the planetary orbits and bulk properties were inferred; we provide both a qualitative discussion (\S \ref{sec:details:quant}) and a more detailed description of the photometric-dynamical model (\S \ref{sec:photo}), model parameters, and the posterior estimation technique.  In \S \ref{sec:planets}, we derive constraints on the composition of the planets based on their bulk densities and thermal histories.  We discuss the short-term evolution in \S \ref{sec:short} and long-term stability of Kepler-36 in \S \ref{sec:stab}.  \deletion{Finally, w}\addition{W}e compare and contrast Kepler-36 with other known planetary systems \addition{in \S~\ref{sec:comp}. Finally, in \S~\ref{sec:avail}, we describe how to obtain the data used in this analysis and describe attached files containing additional information relevant to this work (including a sample of our MCMC results).}}

 \section{Stellar properties of Kepler-36} \label{sec:star}
\subsection{Spectroscopy} \label{sec:star:spec}
\subsubsection{Observations}

Spectroscopic observations to determine the stellar characteristics of Kepler-36 were conducted
independently at two observatories. Two spectra were acquired using the Tull Coud\'{e} Spectrograph on
the 2.7 m the Harlan J. Smith Telescope at the McDonald Observatory Texas on 29 March 2010 and 8 April
2010. The two spectra were shifted and co-added to provide higher SNR for the spectroscopic analysis. One
spectrum was acquired using the HIRES spectrometer on the Keck I 10 m telescope on 11 April 2011.

\subsubsection{Determination of photospheric stellar parameters}

We used the Stellar Parameter Classification (SPC) method \cite{Buchhave2012}, to derive the stellar
atmosphere parameters from the observed spectra. SPC cross-correlates the observed spectrum against a
grid of synthetic spectra drawn from a library calculated by John Laird using Kurucz models \cite{Kurucz1992}. 
The synthetic spectra cover a window of 300 \AA  ~centered near the gravity-sensitive Mg b features and has a
spacing of 250 K in effective temperature, 0.5 dex in gravity, 0.5 dex in metallicity and a progressive spacing in rotational velocity starting at 1 km s$^{-1}$ up to 20 km s$^{-1}$. To derive the precise stellar parameters between the grid points, the normalized cross-correlation
peaks were fitted with a three dimensional polynomial as a function of effective temperature, surface
gravity and metallicity. This procedure was carried out for different rotational velocities and the final stellar
parameters were determined by a weighted mean of the values from the spectral orders covered by the
library.

Initial values for the effective temperature and metallicity were determined and used as input in the
asteroseismic analysis detailed in Section \ref{sec:astero} to estimate the surface gravity of the host star. SPC was
then re-run, fixing the surface gravity to the value from the asteroseismic analysis (log(g) = 4.045). The final
stellar parameters reported here, effective temperature $T_{\rm eff} = 5911 \pm 66 K$, metallicity [m/H] = $-0.20 \pm
0.06$ and $v \sin i = 4.9 \pm 1.0$ km$^{-1}$, are the mean values derived from the co-added McDonald spectrum.  The HIRES spectrum was provided
late to this analysis, but its parameters agree with those derived from the McDonald spectrum.

 \subsection{Asteroseismic analysis of Kepler-36}
 \label{sec:astero}

The asteroseismic analysis of Kepler-36 was performed by the Kepler
Asteroseismic Science Operations Centre (KASOC) team.

 \subsubsection{Estimation of asteroseismic parameters}
 \label{sec:extract}

\begin{figure*}
\center
\includegraphics[width=5.in]{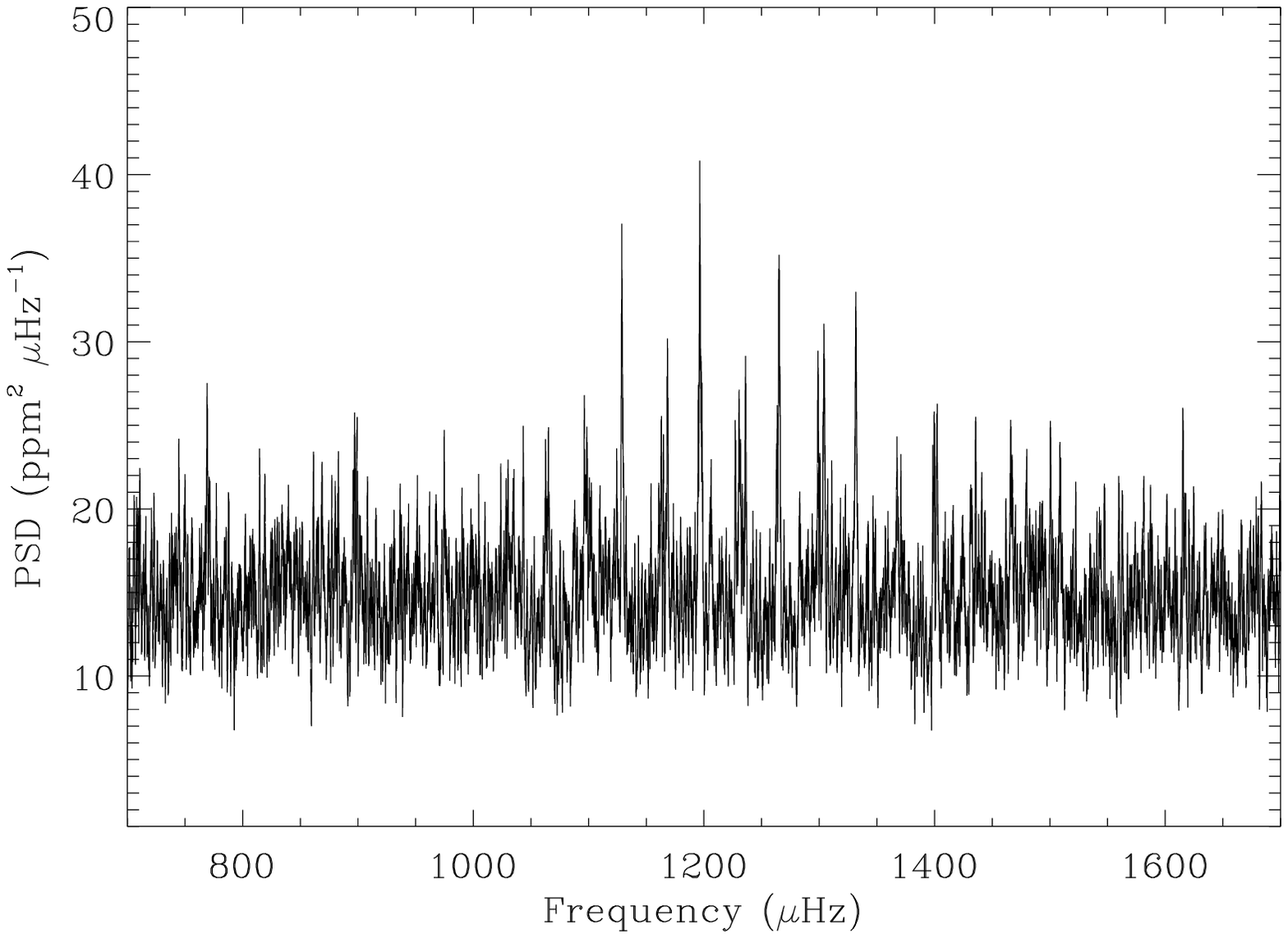}
 \caption{Frequency-power spectrum of Kepler-36, showing a pattern of nearly equally-spaced 
   overtones of solar-like oscillations.}
\label{fig:powspec}
\end{figure*}

The analysis was based upon 15\,months of {\it Kepler} short-cadence (1-minute sampled)
data, collected between 2010 March 2 and 2011 June 26.
Fig.~\ref{fig:powspec} plots the frequency-power spectrum of the
light curve (the planetary transit signals having first been removed).
It shows a clear pattern of peaks due to solar-like oscillations that
are acoustic (pressure, or p) modes of high radial order, $n$. The
observed power in the oscillations is modulated in frequency by an
envelope that has an approximately Gaussian shape. The frequency of
maximum oscillation power, $\nu_{\rm max}$, has been shown to scale to
good approximation as $gT_{\rm eff}^{-1/2}$ \cite{1991ApJ...368..599B,1995A&A...293...87K}, %(Brown et al. 1991; Kjeldsen \& Bedding 1995), 
where $g$ is the surface gravity and
$T_{\rm eff}$ is the effective temperature of the star.  The most
obvious spacings in the spectrum are the large frequency separations,
$\Delta\nu$, between consecutive overtones $n$ of the same spherical
angular degree, $l$. These large separations scale to very good
approximation as $\left< \rho \right>^{1/2}$, $\left< \rho \right>
\propto M/R^3$ being the mean density of the star with mass $M$ and
surface radius $R$ \cite{1993ASPC...42..347C}. %(e.g. see Christensen-Dalsgaard 1993).

We used four independent analysis codes to obtain estimates of the
average large separation, $\left< \Delta\nu \right>$, and $\nu_{\rm
  max}$, using automated analysis tools that have been developed, and
extensively tested \cite{2010ApJ...713L.164C,2010MNRAS.402.2049H,2009CoAst.160...74H,2011MNRAS.415.3539V}  %(Christensen-Dalsgaard et al. 2010; Hekker et al. 2010; Huber et al. 2009; Verner et al. 2011) 
for application to
{\it Kepler} data \cite{Chaplin3322011}. %(Chaplin et al. 2011). 
A final value of each
parameter was selected by taking the individual estimate that lay
closest to the average over all teams. The uncertainty on the final
value was given by adding (in quadrature) the uncertainty on the
chosen estimate and the standard deviation over all teams. Excellent
agreement was found between the results. The final values for $\left<
\Delta\nu \right>$ and $\nu_{\rm max}$ were $67.9 \pm 1.2\,\rm \mu Hz$
and $1250 \pm 44\,\rm \mu Hz$, respectively.

Use of individual frequencies increases the information content
provided by the seismic data for making inference on the stellar
properties. We therefore also applied ``peak bagging'' to the
frequency-power spectrum to extract estimates of the mode
frequencies. Two codes were applied, one using a pseudo-global
approach with maximum likelihood estimation \cite{2009ApJ...694..144F};
and another which performed a global fitting using Markov chain Monte
Carlo (MCMC) \cite{2011A&A...527A..56H}. %(Handberg \& Campante 2011). 
As per the average seismic
parameters, we found excellent agreement in the fitted
frequencies between the two approaches. The frequencies, $\nu^{\rm obs}_{nl}$, and their
associated uncertainties, $\sigma^{\rm obs}_{nl}$, are listed in
Table~\ref{tab:freqs}. They were taken from the MCMC analysis, which
provides more conservative frequency uncertainties.

%%%%%%%%%%%%%%%%%%%%%%%%%%%%%%%%%%%%%%%%%%%%%%%%%%%%%%%%%%%%%%%%%%%%%%%%%

%%%%%%%%%%%%%%%%%%%%%%%%%%%%%%%%%%%%%%%%%%%%%%%%%%%%%%%%%%%%%%%%%%%%%%%%%

%%%%%%%%%%%%%%%%%%%%%%%%%%%%%%%%%%%%%%%%%%%%%%%%%%%%%%%%%%%%%%%%%%%%%%%%%

\begin{table}
\begin{center}
\caption{Estimated frequencies $\nu^{\rm obs}_{nl}$ and uncertainties
  $\sigma^{\rm obs}_{nl}$ (68\,\% credible region) of Kepler-36 (in $\rm
  \mu Hz$).}
\begin{tabular}{cccc}
   &      &      &      \\
\hline
   &      &      &      \\
$n$& $l=0$& $l=1$& $l=2$\\
   &                       &                        &                       \\
14 &                        & $1063.98^{+0.96}_{-1.03}$ & $1095.98^{+1.54}_{-5.08}$ \\
   &                       &                        &                       \\
15 & $1099.15^{+1.59}_{-1.22}$ & $1129.06^{+0.28}_{-0.28}$ & $1162.36^{+0.67}_{-0.66}$ \\
   &                       &                        &                       \\
16 & $1168.35^{+0.35}_{-0.44}$ & $1196.75^{+0.25}_{-0.25}$ & $1230.63^{+0.56}_{-0.95}$ \\
   &                       &                        &                       \\
17 & $1235.86^{+0.30}_{-0.56}$ & $1264.75^{+0.33}_{-0.38}$ & $1298.86^{+0.59}_{-0.65}$ \\
   &                       &                        &                       \\
18 & $1304.01^{+0.29}_{-0.28}$ & $1331.62^{+0.39}_{-0.43}$ &                        \\
   &                       &                        &                       \\
19 &                        & $1400.23^{+0.93}_{-0.71}$ &                       \\
   &                       &                        &                       \\
20 &                        & $1466.60^{+1.28}_{-0.88}$ &                       \\
   &                       &                        &                       \\
\hline
\end{tabular}
\end{center}
\label{tab:freqs}
\end{table}

%%%%%%%%%%%%%%%%%%%%%%%%%%%%%%%%%%%%%%%%%%%%%%%%%%%%%%%%%%%%%%%%%%%%%%%%%

\subsection{Estimation of stellar properties}
 \label{sec:prop}

The analysis was divided into two parts.

 \subsubsection{Grid-based modeling with average asteroseismic parameters}
 \label{sec:grid}

In the first part we used a grid-based approach, in which properties
were determined by searching among a grid of stellar evolutionary
models to get a best fit for the input parameters, which were $\left<
\Delta\nu \right>$, $\nu_{\rm max}$, and the spectroscopically
estimated $T_{\rm eff}$ and [Fe/H] of the star. Descriptions of the
grid-based pipelines used in the analysis may be found in 
\cite{2009ApJ...700.1589S, 2010ApJ...710.1596B, 2010ApJ...725.2176Q, 2011ApJ...730...63G}. %Stello et al. (2009), Basu et al. (2010), Quirion et al. (2010) and Gai et al. (2011). 
An important output of this first part was not only a
guideline set of stellar properties -- which could be used as the
starting points for detailed modeling of the individual frequencies
in the second part of the analysis -- but also iterated, and improved
estimates of $T_{\rm eff}$ and [Fe/H]. Initial values of the
spectroscopic data were used together with the average seismic
properties to estimate log\,$g$. The spectroscopic analysis was then
repeated with log\,$g$ fixed at this value, to yield the revised (and
final) values of $T_{\rm eff} = 5911 \pm 66\,\rm K$ and [Fe/H]$=-0.20
\pm 0.06$. Convergence of the inferred properties (to within the
estimated uncertainties) was achieved after this one iteration.

 \subsubsection{Modeling with individual mode frequencies}
 \label{sec:freq}

In the second and final part of the analysis, the individual
frequencies $\nu^{\rm obs}_{nl}$ were used as the seismic inputs to a
detailed modeling performed by four members of the KASOC team (TM,
JCD, AM and SB). The spectroscopic inputs were also used, as per the
grid-pipeline analysis. Estimated stellar properties from the first,
grid-based part were used either as starting guesses or as a guideline
check for initial results. Each analysis sought to minimize a standard
$\chi^2$ metric. A separate value of $\chi^2$ was calculated for the
asteroseismic and spectroscopic constraints, and these values are
averaged for the final quality metric.

TM used the Asteroseismic Modeling Portal (AMP), a web-based tool tied
to TeraGrid computing resources that uses a parallel genetic algorithm
\cite{2003JCoPh.185..176M} %(Metcalfe \& Charbonneau 2003) 
to optimize, in an automated manner,
the match to observational data \cite{2009ApJ...699..373M,2009gcew.procE...1W}.%  (see {\it Metcalfe et al. 2009, Woitaszek et al. 2009} for more details). 
AMP employs the Aarhus stellar
evolution code ASTEC \cite{2008Ap&SS.316...13C} %(Christensen-Dalsgaard 2008a) 
and adiabatic
pulsation code ADIPLS \cite{2008Ap&SS.316..113C}. %(Christensen-Dalsgaard 2008b).  
Models were made
using the OPAL 2005 equation of state and the most recent OPAL
opacities supplemented by %Alexander \& Ferguson (1994) 
opacities at
low temperature  \cite{1994ApJ...437..879A}, nuclear reaction rates from \cite{1992RvMP...64..885B}, %Bahcall \& Pinsonneault (1992), 
and helium diffusion and settling following \cite{1993ASPC...40..246M}. %Michaud \& Proffitt (1993). 
Convection was treated with standard mixing-length
theory without overshooting \cite{1958ZA.....46..108B}. %(B{\"o}hm-Vitense 1958).

Each AMP model evaluation involves the computation of a stellar
evolution track from the zero-age main sequence (ZAMS) through a
mass-dependent number of internal time steps, terminating prior to the
beginning of the red giant stage. Exploiting the fact that $\left<
\Delta\nu \right>$ is a monotonically decreasing function of age \cite{2009ApJ...699..373M}, %(see Metcalfe et al. 2009, and references therein), 
the asteroseismic age
is optimized along each evolutionary track using a binary decision
tree. The frequencies of the resulting model are then corrected for
surface effects following the prescription of \cite{2008ApJ...683L.175K}. %Kjeldsen et al. (2008).
The optimal model is then subjected to a local analysis that uses
singular value decomposition (SVD) to quantify the uncertainties of
the final model parameters \cite{2007ApJ...659..616C}. %(see Creevey et al. 2007).

JCD applied a fitting technique that has been used for the analysis of
the Hubble observations of HD 17156 \cite{2011ApJ...726....2G}, %(Gilliland et al., 2011), 
and {\it Kepler} observations of the {\it Kepler} exoplanet host stars
HAT-P-7 \cite{2010ApJ...713L.164C} %(Christensen-Dalsgaard et al., 2010) 
and Kepler-10 \cite{2011ApJ...729...27B}.  %(Batalha et al., 2011). 
The stellar modeling was carried out with the
ASTEC code \cite{2008Ap&SS.316...13C}. %(Christensen-Dalsgaard, 2008a).  
The calculations used the
OPAL equation of state tables \cite{1996ApJ...456..902R} %(see Rogers et al. 1996) 
and OPAL
opacities at temperatures above $10^4\,\rm K$ \cite{1996ApJ...464..943I}; %(Iglesias \& Rogers 1996);
at lower temperature the  %Ferguson et al. (2005) 
opacities \cite{2005ApJ...623..585F} were
used. Nuclear reactions were calculated using the NACRE parameters
\cite{1999NuPhA.656....3A}. %(Angulo et al., 1999). 
Convection was treated using the \cite{1958ZA.....46..108B} %B\"ohm-Vitense (1958) 
mixing-length formulation. Frequencies were computed for the
models using ADIPLS \cite{2008Ap&SS.316..113C}. %(Christensen-Dalsgaard, 2008b). 
The prescription in \cite{2008ApJ...683L.175K} %Kjeldsen et al. (2008) 
was again applied in an attempt to deal with
the surface term.

For each evolutionary sequence in the grid of ASTEC models, the model
${\cal M}'_{\rm min}$ whose surface-corrected frequencies provided the
best $\chi^2$ match to the observations was selected. The best match
was obtained from application of homology scaling, i.e., it was
assumed that frequencies in the vicinity of ${\cal M}'_{\rm min}$
could be obtained as $r\nu_{nl}({\cal M}'_{\rm min})$, where
$r=[R/R({\cal M}'_{\rm min})]^{-1.5}$, $R$ being the surface radius of
the model. A best-fitting model was then determined by minimizing the
sum $\sum \left( \nu^{\rm obs}_{nl} - r\nu_{nl}({\cal M}'_{\rm min})
\right)^2/ \left(\sigma^{\rm obs}_{nl} \right)^2$ over all observed
modes, as a function of $r$. The resulting value of $r$
defines an estimate of the radius of the best-fitting model along the
given sequence. The other properties of this best-fitting model are
then determined by linear interpolation in $R$, to the radius of the best-fitting model.
Statistical analysis of the ensemble of best-fitting properties from
all evolutionary sequences then yielded the final stellar properties,
and their uncertainties. A Monte Carlo simulation, involving 100
realizations of the observed frequencies with the addition of normally
distributed errors having the inferred standard deviations, showed
that the observational errors on the frequencies make a modest
contribution to the errors in the inferred stellar properties.

AM used stellar models computed with the CLES code \cite{2008Ap&SS.316..149S}, %(Scuflaire et al. 2008a), 
from the pre-main sequence up to the sub-giant branch. For
each evolutionary track adiabatic oscillations were computed for about
120 main-sequence and sub-giant models with the LOSC code \cite{2008Ap&SS.316...83S} %(Scuflaire et al. 2008b).  
Grids of models both without and with diffusion and
settling of helium and heavy elements (see \cite{2008Ap&SS.316..149S} were
considered.  The opacity tables are those of OPAL96 \cite{1996ApJ...464..943I} %(Iglesias \& Rogers 1996) 
complemented at $T<10^4\,\rm K$ with the opacities of
\cite{2005ApJ...623..585F}. % Alexander \& Ferguson (2005). 
The metal mixture used in the opacity
tables was the solar one, as per \cite{1993PhST...47..133G}. %Grevesse \& Noels (1993). 
The nuclear
energy generation routines are based on the cross-sections by NACRE
\cite{1999NuPhA.656....3A}, %(Angulo et al. 1999), 
with the OPAL equation of state \cite{2002ApJ...576.1064R}. %(Rogers \& Nayfonov 2002). 
Convection transport was treated with the classical
mixing length approximation of \cite{1958ZA.....46..108B}. %B\"ohm-Vitense (1958).

The analysis was performed in a similar manner to JCD, including
treatment of the surface term using the  %Kjeldsen et al. 
prescription \cite{2008ApJ...683L.175K},
and interpolation in $R$ close to the $\chi^2$ minimum to find the
best-fitting properties.

SB made use of the Yale stellar evolution code, YREC \cite{2008Ap&SS.316...31D} %(Demarque et al. 2008) 
to model the star.  The input physics included the OPAL
equation of state tables of \cite{2002ApJ...576.1064R}, %Rogers \& Nayfonov (2002), 
and OPAL
high-temperature opacities \cite{1996ApJ...464..943I} %(Iglesias \& Rogers 1996) 
supplemented with
low -temperature opacities from \cite{2005ApJ...623..585F}. %Ferguson et al.~(2005).  
All nuclear
reaction rates were from \cite{1998RvMP...70.1265A}, %Adelberger et al.~(1998) 
except for the rate
of the $^{14}{\rm N}(p,\gamma)^{15}{\rm O}$ reaction, which was fixed
at the value of \cite{2004PhLB..591...61F}. %Formicola et al.~(2004). 
Models were constructed for
two values of core overshoot, 0 and 0.2$H_p$. Two families of models
were constructed, one that included the diffusion and settling of
helium and heavy elements as per the the formulation of \cite{1994ApJ...421..828T}, %Thoul et al.~(1994), 
and one that did not include any diffusion and settling.

YREC was used in an iterative mode. In this mode the final $T_{\rm
  eff}$ and radius for a model of a given mass and metallicity was
specified, and for a given mixing length parameter $\alpha$ the code
iterated over the initial helium abundance $Y_0$ until a model with
the specified $T_{\rm eff}$ and radius was found.  Note that this is
similar to the construction of standard solar models, though in that
case the iteration is performed over both the mixing length parameter
and $Y_0$ (since the solar age is a known constraint).  Since the age
of Kepler-36 is not known independently, the iteration over $Y_0$ was
done for many different values of the mixing length parameter.  All
solutions for which the initial helium abundance was less than the
primordial helium abundance, $Y_{\rm p}$ were rejected. $Y_{\rm p}$
was assumed to be $0.245$.

The surface term correction was handled in a manner that is slightly
different from the other analyses. The first step was the construction
of a standard solar model with exactly the same physics as that used
to model Kepler-36. This yielded the set $\nu_{nl\odot}$ of solar model
frequencies. These were then used to estimate a set of ``surface
term'' frequency offsets, $\delta\nu_{nl\odot}$, for the Sun by
computing differences between the solar model frequencies and the
solar low-degree frequencies observed by the Birmingham Solar
Oscillations Network (BiSON), as in \cite{2009ApJ...699.1403B}. %(as listed in Basu et al.~2009).

Then, for each stellar model ${\cal M'}$ under consideration,
$\nu_{nl\odot}$ and $\delta\nu_{nl\odot}$ were scaled to the mass and
radius of ${\cal M'}$ using the homology scaling $r = \left<
\Delta\nu({\cal M'}) \right> / \left< \Delta\nu_{nl,\odot}
\right>$. The resulting $r\nu_{nl\odot}$-$r\delta\nu_{nl\odot}$
relation was then used to correct the stellar model for the surface
term. Using a least squares minimization a factor $\beta$ was selected
so as to minimize $\sum \left(\nu^{\rm obs}_{nl}-\nu^{\rm corr}_{nl}
\right)^2/ \left(\sigma^{\rm obs}_{nl} \right)^2$ over all observed
modes, where $\nu^{\rm corr}_{nl}=\nu_{nl}^{\cal M'}+
\beta\;r\delta\nu_{nl\odot}$, with $r\delta\nu_{nl\odot}$ evaluated at
$r\nu_{nl\odot}=\nu^{\rm obs}_{nl}$.

 \subsubsection{Results on stellar properties}
 \label{sec:res}

Kepler-36 turned out to be a non-trivial star to model. The star is
quite evolved, and a fraction of the models that can potentially
provide a reasonable match to the observations show mixed $l=2$ modes
in the frequency range where estimated frequencies are provided,
thereby giving the potential to complicate the analysis \cite{2010A&A...515A..87D,2010ApJ...723.1583M}. %(Deheuvels et al. 2010; Metcalfe et al. 2010). 
Since none of the observed
frequencies showed evidence of strong mixing, model frequencies
identified as those with the highest inertia, which did not satisfy
asymptotic behaviour, e.g. \cite{1980ApJS...43..469T,1986hmps.conf..117G}, %(e.g., Tassoul 1980; Gough 1986), 
were eliminated
from the analysis. Some modifications were also made to the \cite{2008ApJ...683L.175K} %Kjeldsen et al. (2008) 
surface-term prescription. This prescription was
developed for application to radial modes: However, since for Kepler-36
the observed radial-mode data in Table~\ref{tab:freqs} does not extend
to the highest frequencies, JCD adapted the method so it could be
applied to the non-radial modes. The correction was based on an
estimate of the large separation $\left< \Delta \nu \right>$ and an
average frequency obtained by fitting to all the observed modes.

In spite of the above potential complications, differences in the
treatment of the surface term, and differences in the physics inputs,
there was excellent agreement in the properties returned by all four analyses. For each of $M$, $R$, log\,$g$, mean density $\left< \rho
\right>$, and age $\tau$, the standard deviation of the four estimates
was significantly lower than the median formal uncertainty of the
property.

Our final properties are those of the modeler whose estimates lay
closest to the median (and mean) over all modelers.  For the
uncertainties we took the chosen modeler's uncertainty for each
property and added (in quadrature) the standard deviation of the
property from the different modeling results. We find $M = 1.071 \pm
0.043\,\rm M_{\odot}$, $R = 1.626 \pm 0.019\,\rm R_{\odot}$, $\left<
\rho \right> = 0.3508 \pm 0.0056\,\rm g\,cm^{-3}$, $\log g = 4.045 \pm
0.009\,\rm dex$ and $\tau= 6.8 \pm 1.0\,\rm Gyr$. (We add that the
initial, grid-modeling-based estimates agreed to within
uncertainties) Kepler-36 has a luminosity 2.9 times that of the Sun and an approximate distance of 470 pc.

\subsection{Photometric determination of the stellar rotation period?}

\begin{figure}
\centering
\includegraphics[width=6in]{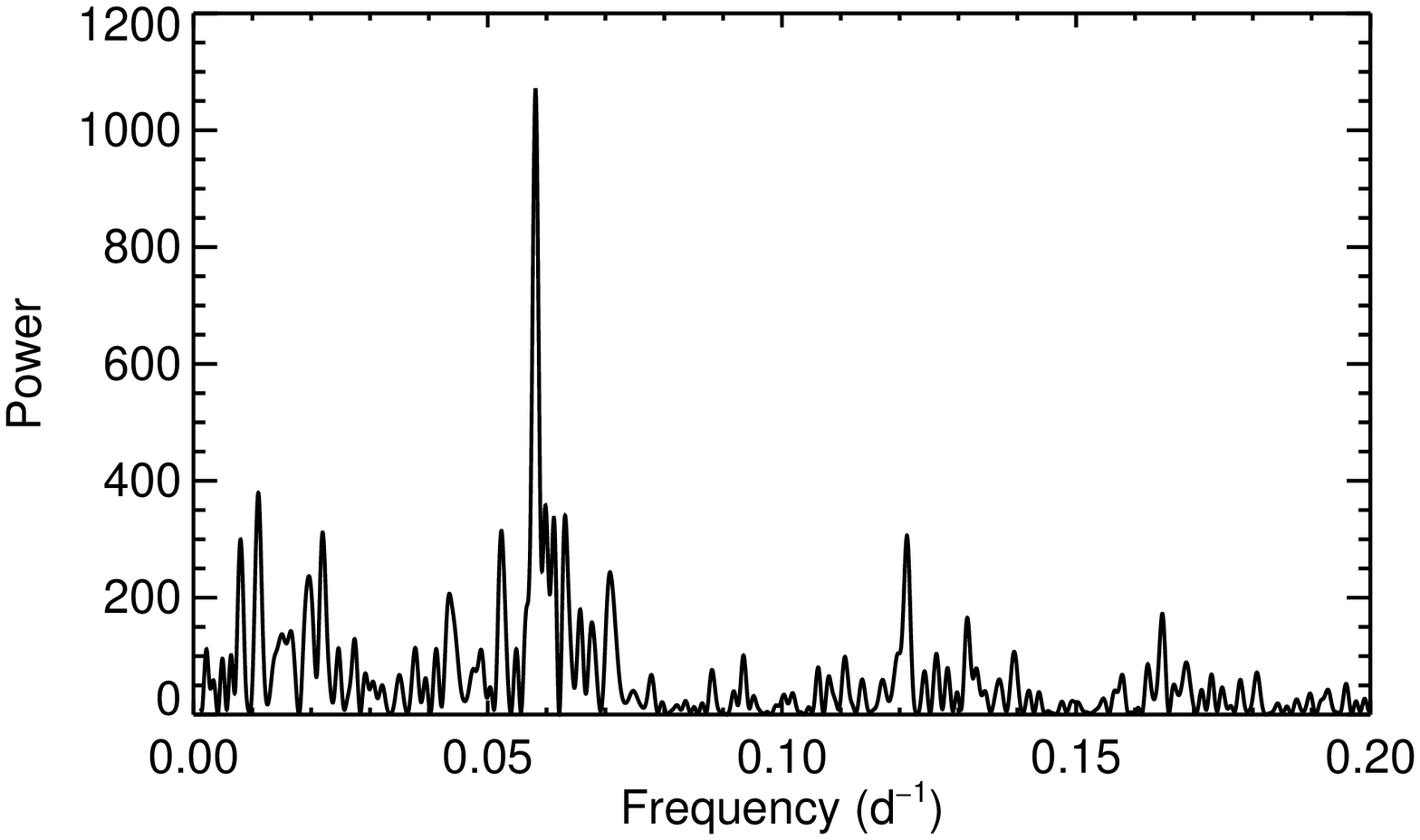}
\caption{Periodogram of the {\it Kepler} photometry at low frequencies demonstrating a strong peak at 
$0.058$ d$^{-1}$ or 17.20 days.  This may represent the rotational period of Kepler-36; it agrees with the 
maximum rotational period determined from $v\sin i$ and the radius of Kepler-36. \label{fig:ls}}
\end{figure}

The full {\it Kepler} photometry demonstrates a significant modulation with a period of $17.20\pm0.2$ days 
(Figure \ref{fig:ls}).  This period corresponds to the rotational period of Kepler-36 if this modulation is due to the ``transit'' of photospheric surface features.   
In fact, this measured period agrees well with the maximum rotational period; we find $P_{\rm rot, max} = 16.8 \pm 4.3$ 
days assuming the results from above of $v \sin i_\star = 4.9\pm 1.0$ km/s and $R_\star = 1.626 \pm 0.019$.  A $17.20$ day 
rotational period is typical \cite{deMedeiros3171997} for subgiant stars with the same $B-V$ color as Kepler-36 (estimated to be $B-V \approx 0.55$).  We note that this period is close to the orbital period of Planet c (at 16.28 days).  However, it is extremely unlikely that the star has been tidally synchronized with c given its mass and distance.  The correspondence is likely coincidental.

\section{Constraints on Planetary Orbits, Masses and Radii} \label{sec:details}

In this section we describe in detail how we derive the properties of the planets and the orbits
from the photometric data.

\subsection{Qualitative Considerations} \label{sec:details:quant}

\begin{figure}
\centering
\includegraphics[width=6.in]{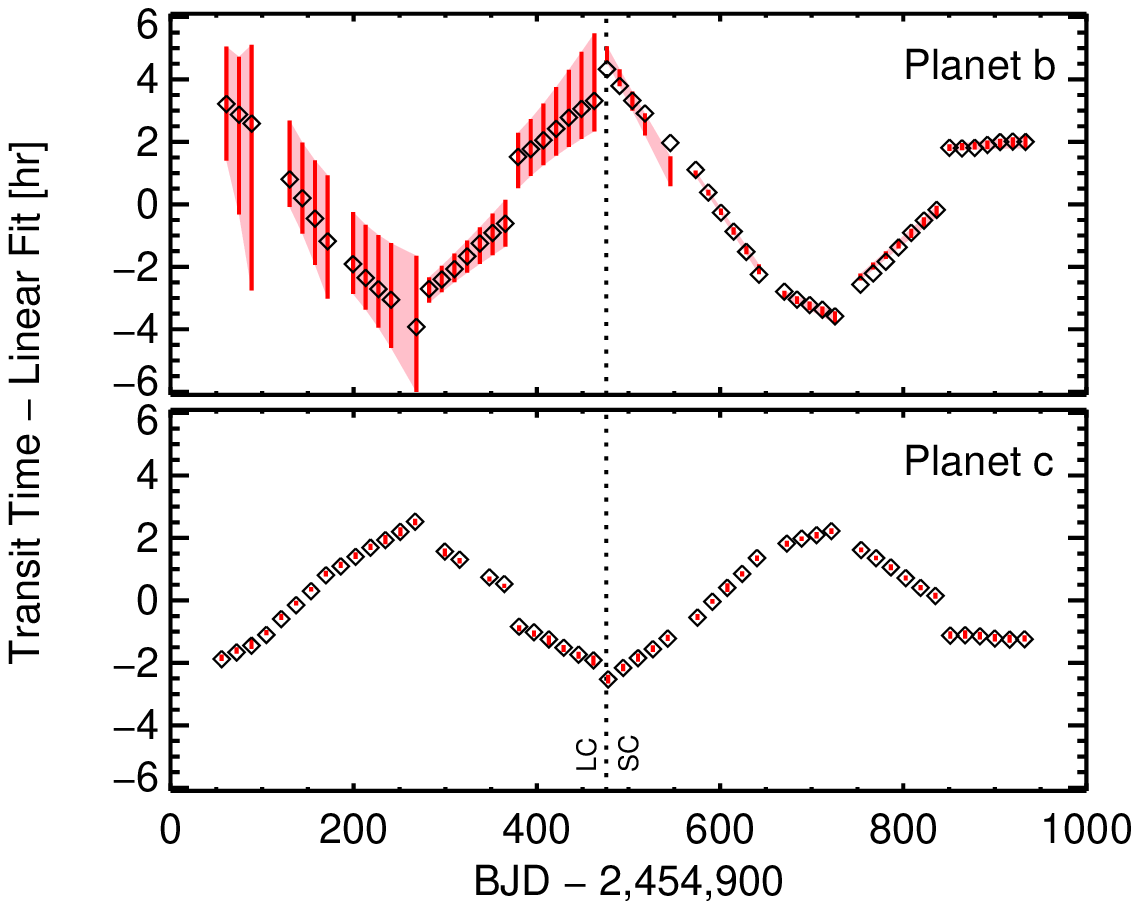}
\caption{Transit times of the two planets as predicted from a fit assuming a linear ephemeris between
each conjunction for Planet b and individual epoch transit times for Planet c. The error bars represent the uncertainty on each transit
time.  Due to the smaller transit depth for Planet b, the uncertainties are
much larger than for planet c.  The dotted vertical line indicates the time at which
short cadence data began to be collected for Kepler-36;  the
short cadence data yield much more precise times of transit. \addition{The open diamonds are the best-fit times according to the photodynamical model (\S~\ref{sec:photo}).  The typical uncertainty in the predicted times are 13 minutes and 3 minute for the transits of planet b and planet c, respectively.}  \label{fig:ttv}}
\end{figure}

Mass ratios between the planets and the star may be quickly estimated based on the qualitative behavior of the variations of the times of transit relative to a constant period model [i.e., the functional form of the transit timing variations (TTVs); see Figure \ref{fig:ttv}, Tables \ref{tab:tabmtb} and \ref{tab:tabmtc}], as described as follows.

 \begin{table}[h]
\centering
\begin{tabular}{lcc}
\hline
Epochs & Ephemeris, $T_0$  & Period, $P$ (days) \\
&  (BJD-2,454,900) & \\ \hline
0,
1,
2,
&
$ 60.9771\pm  0.0760$
& $ 13.8073\pm  0.0725$ \\
5,
6,
7,
8,
&
$130.1465\pm  0.0576$
& $ 13.8174\pm  0.0196$ \\
10,
11,
12,
13,
15,
&
$199.2772\pm  0.0546$
& $ 13.8310\pm  0.0145$ \\
16,
17,
18,
19,
20,
21,
22,
&
$282.3275\pm  0.0169$
& $ 13.8648\pm  0.0044$ \\
23,
24,
25,
26,
27,
28,
29,
&
$379.4499\pm  0.0368$
& $ 13.8673\pm  0.0090$ \\
30,
31,
32,
33,
35,
&
$476.5412\pm  0.0107$
& $ 13.8187\pm  0.0033$ \\
37,
38,
39,
40,
41,
42,
&
$573.3318\pm  0.0037$
& $ 13.8244\pm  0.0010$ \\
44,
45,
46,
47,
48,
&
$670.1201\pm  0.0046$
& $ 13.8423\pm  0.0016$ \\
50,
51,
52,
53,
54,
55,
56,
&
$753.2429\pm  0.0044$
& $ 13.8645\pm  0.0010$ \\
57,
58,
59,
60,
61,
62,
63,
&
$850.3652\pm  0.0044$
& $ 13.8512\pm  0.0012$ \\
 \hline
\end{tabular}
\caption{Linear ephemerides for Planet b, measured independently from the {\it Kepler} photometry.  The ephemeris listed is that for the first epoch in the set under ``Epochs.''  \label{tab:tabmtb}}
\end{table}

The planetary mass ratio can be estimated assuming conservation of
energy.  Since the planets interact gravitationally, if they do not
experience a close approach, they each maintain nearly Keplerian
orbits.  This suggests that 
\begin{eqnarray}
	E_{\rm orbital} &\approx& E_b + E_c \approx -{GM_\star M_b\over 2a_b} -{GM_\star M_c\over 2a_c} \approx  {\rm const}.
\end{eqnarray}
\addition{This equation neglects the interaction energy of the two planets
as well as the the planet masses relative to the star in computing the orbital periods.}
Since $a_b/a_c \approx (P_b/P_c)^{2/3}$,
\begin{eqnarray}
	{M_b \over P_b^{2/3}} &\approx& -{M_c \over P_c^{2/3}}+{\rm const}.  
\end{eqnarray}
Perturbing
this equation gives
\begin{eqnarray}
	 {M_c \over M_b} & \approx & -{\delta P_b \over \delta P_c} \left({P_c^{5/3}\over P_b^{5/3}}\right).
\end{eqnarray}
\addition{This equation assumes that the energy exchange between the planets is small compared to their orbital energies (i.e.\ that there are no close encounters).}
Investigating Figure 2, $\delta P_b \approx 8.5$ hr and $\delta P_c \approx 5.5$ hr,
while $P_c/P_b \approx 7/6 \approx 1.17$, so $M_c/M_b \approx 2$. 

The ratio of the sum of the planet masses to the star can be derived
from the total amplitude of the TTVs and the
period of circulation \cite{Agol3592005}.  The best-fit periods are $P_b=13.85$ days 
and $P_c=16.23$ days and have a proximity to resonance of $\epsilon = 
1-(j+1)P_b/(jP_c) \approx 0.005$ where $j=6$ in this case.

This distance from resonance causes the conjunctions to
drift in inertial space over $N_{conj} \approx  j^{-2} \epsilon^{-1}\approx  5$ 
conjunctions.
As can be seen in Figure 2, it takes $\approx 5$ conjunctions (e.g. from 150
to 600 days when the TTV traces cross one another in the same
direction) to complete a TTV cycle.

The semi-amplitude of the sum of the TTVs is
$\delta t \approx 7$ hours.  Using equation 29 from \cite{Agol3592005}, this gives a 
total mass ratio of 
\begin{eqnarray}
	\mu \equiv \frac{M_b+M_c}{M_\star} &=& \sqrt{\frac{\delta t~\epsilon^3}{P_b }} \approx 5 \times 10^{-5}.
\end{eqnarray}
\addition{This equation assumes that period variations dominate the transit timing variations (which can be seen in the riverplot as the jumps at conjunctions are smaller than the TTV due to changes in period between conjunctions).  It also assumes plane-parallel planets of much smaller mass than the star, and neglects some factors that are of order unity.}

Resonance is expected for $\epsilon < j^{1/3}\mu^{2/3} = 0.0024$,
which is not satisfied for the $6:7$ commensurability.  \addition{The best-fit solution's (described in \S~\ref{sec:best}) resonant argument does not librate.}

\begin{longtable}{lc}
%\centering
%\begin{tabular}{ll}
\caption[]{Planet c mid-transit times, measured independently from the {\it Kepler} photometry. \label{tab:tabmtc}} \\
Epoch & Mid-transit time \\
&(BJD - 2,454,900) \\ \hline
\endfirsthead
\caption[]{continued...} \\
Epoch & Mid-transit time \\
&  (BJD - 2,454,900) \\ \hline
\endhead
\hline
0 & $ 55.9135\pm  0.0042$\\
1 & $ 72.1543\pm  0.0041$\\
2 & $ 88.3951\pm  0.0071$\\
3 & $104.6409\pm  0.0034$\\
4 & $120.8930\pm  0.0039$\\
5 & $137.1427\pm  0.0030$\\
6 & $153.3924\pm  0.0028$\\
7 & $169.6447\pm  0.0036$\\
8 & $185.8876\pm  0.0040$\\
9 & $202.1310\pm  0.0043$\\
10 & $218.3740\pm  0.0041$\\
11 & $234.6149\pm  0.0059$\\
12 & $250.8567\pm  0.0058$\\
13 & $267.1010\pm  0.0043$\\
15 & $299.5228\pm  0.0051$\\
16 & $315.7430\pm  0.0035$\\
18 & $348.1813\pm  0.0032$\\
19 & $364.4032\pm  0.0027$\\
20 & $380.5782\pm  0.0039$\\
21 & $396.8023\pm  0.0041$\\
22 & $413.0243\pm  0.0065$\\
23 & $429.2448\pm  0.0046$\\
24 & $445.4671\pm  0.0045$\\
25 & $461.6911\pm  0.0067$\\
26 & $477.8979\pm  0.0056$\\
27 & $494.1446\pm  0.0048$\\
28 & $510.3889\pm  0.0053$\\
29 & $526.6322\pm  0.0044$\\
30 & $542.8778\pm  0.0040$\\
32 & $575.3683\pm  0.0039$\\
33 & $591.6204\pm  0.0030$\\
34 & $607.8700\pm  0.0056$\\
35 & $624.1196\pm  0.0033$\\
36 & $640.3718\pm  0.0035$\\
38 & $672.8535\pm  0.0038$\\
39 & $689.0915\pm  0.0029$\\
40 & $705.3274\pm  0.0041$\\
41 & $721.5641\pm  0.0037$\\
43 & $754.0018\pm  0.0029$\\
44 & $770.2221\pm  0.0028$\\
45 & $786.4410\pm  0.0039$\\
46 & $802.6580\pm  0.0031$\\
47 & $818.8763\pm  0.0033$\\
48 & $835.0969\pm  0.0038$\\
49 & $851.2757\pm  0.0048$\\
50 & $867.5075\pm  0.0055$\\
51 & $883.7375\pm  0.0050$\\
52 & $899.9658\pm  0.0048$\\
53 & $916.1958\pm  0.0051$\\
54 & $932.4275\pm  0.0041$\\
 \hline
%\end{tabular}

\end{longtable}

\subsection{Photometric-Dynamical Model} \label{sec:photo}

We modeled the
{\it Kepler} light curve of Kepler-36 using a dynamical model to predict the
motions of the planets, and a transit model to predict the light curve.  We performed an initial fit to the measured times of transit (given in Tables \ref{tab:tabmtb} and \ref{tab:tabmtc}), using the dynamical model alone, to seed a subsequent fit with the full photometric-dynamical model.

\subsubsection{Description of the model}

The ``photometric-dynamical model'' refers to the model that was used to fit the {\it Kepler} photometry.  This model is equivalent to that described in the analyses of KOI-126 \cite{Carter3312011}, Kepler-16 \cite{Doyle3332011}, Kepler-34 and Kepler-35 \cite{Welsh4812012}.  

The underlying model was a gravitational three-body integration.  This integration utilized a hierarchical (or Jacobian) coordinate system. In this system, ${\bf r_b}$ is the position of Planet b relative to the star, and ${ \bf r_c}$ is the position of Planet c relative to the center of mass of Planet b and the star.  The computations are performed in a Cartesian system, although it is convenient to express ${\bf r_b}$ and ${\bf r_c}$ and their time derivatives in terms of osculating Keplerian orbital elements: instantaneous period, eccentricity, argument of pericenter, inclination, longitude of the ascending node, and time of transit: $P_{b,c}$, $e_{b,c}$, $i_{b,c}$, $\omega_{b,c}$, $\Omega_{b,c}$, $T_{b,c}$, respectively.  We note that these parameters do not necessarily reflect observables in the light curve; the unique three-body effects make these parameters functions of time. The ``time of transit," in particular, does not exactly correspond to an measured transit time; exact correspondence would only be seen if the orbit proceeded in a Keplerian fashion from the reference epoch.

The accelerations of the three bodies are determined from Newton's equations of motion, which depend on ${\bf r_b}$, ${\bf r_c}$ and the masses \cite{Soderhjelm1411984,Mardling5732002}.  For the purpose of reporting the masses and radii in Solar units, we assumed $G M_{\rm Sun} = 2.959122 \times 10^{-4}$ AU$^{3}$ day$^{-2}$ and $R_{\rm Sun} = 0.00465116$ AU.  We used a Bulirsch-Stoer algorithm \cite{Press2002} to integrate the coupled first-order differential equations for $\dot{\bf r}_{b,c}$ and ${\bf r}_{b,c}$.  

We did not determine the spatial coordinates of all three bodies at each observed time.  Instead, to speed computation, we recorded only the sky-plane projected separation between star and planet  and the sky-plane projected speed of planet relative to star at the calculated time of transit (for a given epoch).  The times of transit were solved for numerically as being those times when the projected separation between the star and planet was minimized.  The result of these calculations was a collection of transit times $t^{k}_{i_k}$, impact parameters $b^{k}_{i_k}$ and speeds $v^{k}_{i_k}$ for each planet $k \in \{b,c\}$ and for epochs $i_k \in N_k$ where $N_k$ is the set of observed epoch numbers for planet $k$.  The motion of the planet relative to the star is approximately linear in the sky-plane such that the projected separation as a function of time is, to good approximation,
\begin{eqnarray}
	Z^k_{i_k}(t) = \sqrt{\left[v^k_{i_k}(t-t^k_{i_k})\right]^2+\left(b^k_{i_k}\right)^2}
\end{eqnarray}
for times near (a few transit durations) of the calculated mid-transit time.

The approximate photometric model for the relative stellar flux, $f(t)$, is then defined as
\begin{eqnarray}
	f(t) = 1-\sum_k \sum_{i_k \in N_k} \left\{
		\begin{array}{ll}
			\lambda\left(Z^k_{i_k}(t),R^k,u\right)\left[1+c(t)\right]^{-1} &  -0.5 \leq t-t^k_{i_k} \leq 0.5 \\
			0 & {\rm otherwise}
		\end{array}
		\right.
\end{eqnarray}
where $\lambda(z,r,u)$ is the overlap integral between a limb darkened star of radius $R_\star$ (such that the radial brightness profile is $I(\rho/R_\star)/I(0) = 1-u [1-\sqrt{1-(\rho/R_\star)^2}]$ with linear limb-darkening parameter $u$) whose center is separated by a distance $z$ from a dark, opaque sphere of radius $r$.  The function $c(t)$ is a piecewise constant function giving the extra flux, relative to the stellar flux, that is included in the photometric aperture and is assumed to be constant in each {\it Kepler} quarter.  $\lambda(z,r,u)$ may be computed semi-analytically with available codes \cite{Mandel5802002}.  This photometric model, assuming constant transit velocity, is faster to compute than calculating the positions at each photometric cadence and results in a negligible change in the quality of the model fit to the data compared to exact integration.  This model does not include the ``anomalous'' brightening events that occur when the Planet c occults Planet b during a transit \cite{Ragozzine2010}.  No such events are predicted within the current observations.

The continuous model $f(t)$ is integrated over a 29.4 minutes interval centered on each long cadence sample.  The continuous model is compared ``as is'' to the short cadence observations.

 \subsubsection{Local detrending of {\it Kepler} data}

The {\it Kepler} light curve (``SAP\_FLUX'' from the standard fits product) for Kepler-36, spanning ten Quarters (877 days), is reduced to only those data within 0.5 day of any transit of either planet.  Only long cadence data ($29.4$ minute cadence) are available for the first 406 days of the available {\it Kepler} data.  Short cadence data (1 minute cadence) are available for the remaining 456 days.  Short cadence data are used when available. \addition{Transits are missing only when {\it Kepler} data are unavailable.  Data are missing as a result of observation breaks during quarterly data transfers or spacecraft safe modes.  }

 Each continuous segment of data (being those data observed within 0.5 day) has a local linear correction divided into it.  The parameters of this linear correction are found through an iterative process, as described as follows.  In the first step, we masked the higher signal-to-noise transits of planet c and then performed a robust linear least-squares fit to each continuous segment.  The data, having divided out this correction, were then fit with the photo-dynamical model with a nonlinear fitter (Levenberg-Marquardt).  The best-fit model was then used to identify cadences in transit which were then masked and the fit and division was repeated.  This process was repeated until the corrections converged to a sufficient tolerance.

We also experimented with performing this correction inline with the posterior estimation routine (described in \S \ref{sec:mcmc}) but found negligible differences compared to leaving these fixed according to the converged linear correction found from the process described above.
 
 \subsubsection{Specification of parameters} \label{sec:params}

The reference epoch was chosen to be $t_0 = 2,454,950$ (BJD), near a particular transit (a somewhat arbitrary choice).

The model has 20 adjustable parameters. Two parameters are related to stellar constraints from asteroseismology and are subject to the priors discussed in \S \ref{sec:res}: the stellar density times the gravitational constant, $G \rho_\star$, and the stellar radius, $R_\star$. Two parameters are the mass ratios $q_+ \equiv (M_b+M_c)/M_\star$ and $q_p \equiv M_b/M_c$. Four parameters encode the eccentricities and arguments of pericenter in a nonlinear way  that reduces the complexity of the posterior topology (resulting in  effectively linear correlations in these parameters):
\begin{eqnarray}
	h_- &\equiv & 0.895 \times e_b\cos \omega_b-e_c \cos \omega_c \\
	h_+ & \equiv & e_b\cos \omega_b+e_c \cos \omega_c\\
	k_- & \equiv & e_b\sin \omega _b- e_c \sin \omega_c \\
	k_+ & \equiv & e_b \sin \omega_b + e_c \sin \omega_c 
\end{eqnarray}

In hindsight, we should have used a different linear combination of parameters, namely
$a_b{\bf e_b} \pm a_c {\bf e_c}$, where ${\bf e_i} = e_i (\cos{\omega_i},\sin{\omega_i})$.
This is due to the fact that the transit timing constrains most strongly the distance at
closest approach and orbital phase of the planets at closest approach. This is a
function of the epicyclic amplitude and phase which is proportional to $a_i{\bf e_i}$ for 
each planet.  The differences are well constrained since these determine the closest
approach for the planets, while the sums are poorly constrained.  The only gain in re-executing our analysis with this parameterization would be more rapid convergence to the target distribution (via the method described in \S \ref{sec:mcmc}).  Having already reached sufficient convergence, we did not pursue this.

The remaining osculating parameters, 7 in total, are the periods $P_b$, $P_c$, the orbital inclinations $i_b$, $i_c$, the times of transit $T_b$, $T_c$ and the difference between the nodal longitudes $\Delta\Omega \equiv \Omega_c-\Omega_b$.  The absolute nodal angle relative to North cannot be determined.

Two more parameters are the relative radii of the planets: $r_b \equiv R_b/R_\star$ and $r_c \equiv R_c/R_\star$.  One parameter, $u$, parameterizes the linear limb darkening law for the star (described above).  The final two parameters describe the width of the probability distribution for the photometric noise of the long cadence and short cadence observations, assumed to be stationary, white and Gaussian-distributed ($\sigma_{\rm LC}$ and $\sigma_{\rm SC}$ -- see the next section).

Additionally, we may specify 9 more parameters describing the function $c(t)$, introduced above, describing the relative extra flux summed in the aperture.  The nine parameters specify the constant extra flux in each Kepler quarter.  There are 10 quarters in total, but the absolute level of this contamination cannot be determined from the photometry alone, but, the levels relative to a single quarter may be specified.  We chose this reference quarter to be quarter 7 having observed in preliminary fits that this quarter's contamination fraction was near the median of the sample of 10 quarters.   The addition of these degrees of freedom did not significantly change the distribution of the 20 primary parameters.

We note that a discrete degeneracy exists between the relative nodal angle,
$\Delta \Omega$, and inclinations, $i_b$ and $i_c$.  The light curve model is invariant under the transformation $(\Delta \Omega,i_b, i_c) \rightarrow 
(-\Delta \Omega, \pi-i_b,\pi-i_c)$.  Consequently, we expect the marginalized posterior distribution of $\Delta \Omega$ to be symmetric about $\Delta \Omega = 0$, and
the distributions of $i_b, i_c$ to be symmetric about $\pi/2$.  

 \subsubsection{Priors and Likelihood} \label{sec:like}

We assumed uniform priors  in 16 of the 20 primary parameters described in the previous section excluding $h_{+,-}$ and $k_{+,-}$.  For these latter four parameters, we enforced uniform priors in eccentricities and arguments of pericenter.  For these priors, the probability density obeys
\begin{eqnarray}
	p( h_{+,-}, k_{+,-}) d h_{+,-}  k_{+,-} &\propto& p(e_{b,c},\omega_{b,c}) \times  \frac{1}{e_b e_c} d_{e_b,e_c} d_{\omega_b, \omega_c} \propto \frac{1}{e_b e_c} d_{e_b,e_c} d_{\omega_b, \omega_c}
\end{eqnarray}

The likelihood ${\cal L}$ of a given set of parameters was taken to be the product of likelihoods based on the photometric data, assumed-Gaussian asteroseismology priors, and the weighting to enforce priors (described above):
\begin{eqnarray}
	{\cal L} &\propto&  \sigma_{\rm LC}^{-N_{\rm LC}} \exp \left[-\sum_i^{N_{\rm LC}} \frac{(\Delta F^{LC}_i)^2}{2 \sigma_{\rm LC}^2} \right] \times  \sigma_{\rm SC}^{-N_{\rm SC}} \exp \left[-\sum_i^{N_{\rm SC}} \frac{(\Delta F^{SC}_i)^2}{2 \sigma_{\rm SC}^2} \right] \\ \nonumber
			&& \times \exp \left[\frac{1}{2}\left(\frac{\Delta G\rho_\star}{\sigma_{G\rho_\star}}\right)^2\right] \times \exp \left[ \frac{1}{2}\left(\frac{\Delta R_\star}{\sigma_{R_\star}}\right)^2\right] \times  \frac{1}{e_b e_c} \nonumber
\end{eqnarray}
where $\Delta F^{\rm LC, SC}_i$ is the $i$th photometric data residual (either long cadence or short cadence), $\Delta G\rho_\star/\sigma_{G\rho_\star}$ and $\Delta R_\star/\sigma_{R_\star}$ are the deviates between the asteroseismic constraints in density and radius, and $\sigma_{\rm LC,SC}$ are the width parameters describing the photometric noise for either long or short cadence data.  These noise parameters have values that are within a fraction of a percent of the root-mean-square normalized flux as calculated using the {\it Kepler} light curve, excluding the data near any transit.

\subsubsection{Best-fit model} \label{sec:best}

We determined a best-fit model by maximizing the likelihood ${\cal L}$ as defined above.  The maximum likelihood solution was found by determining the highest likelihood in a large draw from the posterior as calculated with a Markov Chain Monte Carlo simulation as described in \S \ref{sec:mcmc}.  

\begin{figure}
\centering
\includegraphics[width=5in]{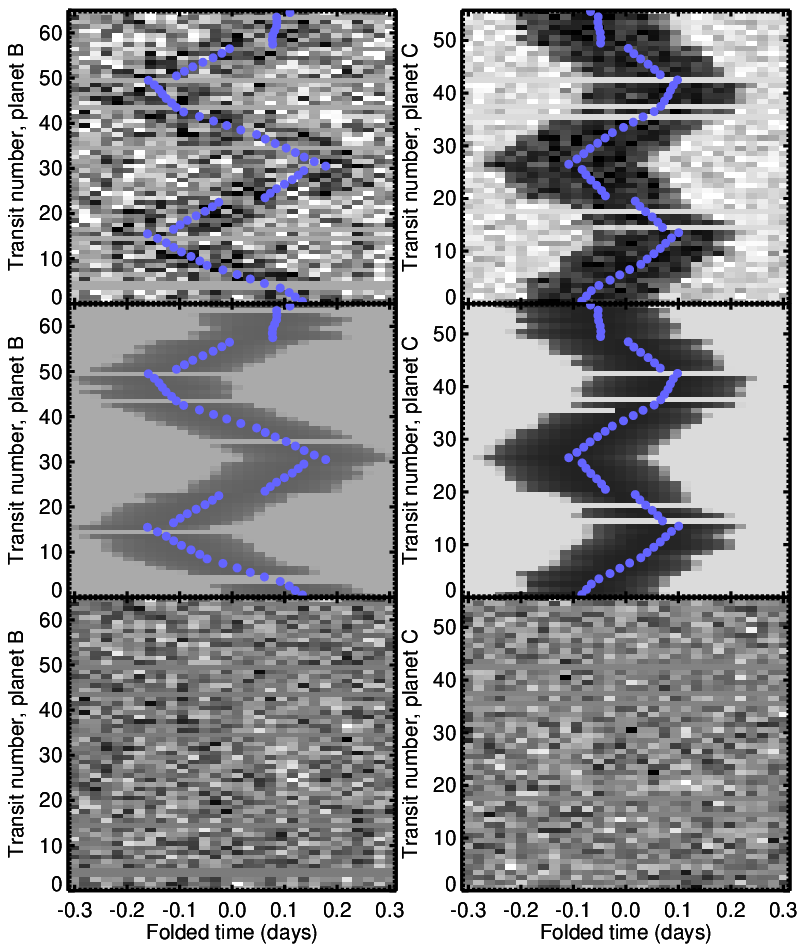}
\caption{Plot of the transits of planet b (left) and planet
c (right).  Top is data; \deletion{bottom}\addition{middle} is best-fit photometric-dynamical model\addition{; bottom is the residuals after subtracting the model}. Refer to the caption of Figure 1 for additional details.
The greyscale ranges from 0.9994 (black) to 1.0001 (white) on a common scale between the \addition{top} four panels. \label{riverplot}}
\end{figure}

Figure \ref{riverplot} shows the data folded on the best-fit period for both planets, as in Figure 2,
as well as the analogous map for the best-fit photometric-dynamical model.   The transit direction has been scaled by the ratio $7:6$ such
that the transits of the planets adjacent to one another correspond to nearly coincident 
times.  The gravitational kicks imparted by the planets on one another are apparent after each conjunction.  The purple dots mark the best-fit time of transit at each epoch; the best-fit times are listed in Tables \ref{tab:tabbfb} and \ref{tab:tabbfc}.  \addition{The best-fit model has $\chi^2 = 99640.7$ for $99718$ degrees of freedom.}  If we ignore the transits of b in the model (or equivalently we let the radius of b be zero) we find $\Delta \chi^2 = 498$; this corresponds to a detection of b at 22 $\sigma$.

\begin{longtable}{lc}
%\centering
%\begin{tabular}{ll}
\caption[]{Planet b mid-transit times from the best-fitting photometric-dynamical model. \label{tab:tabbfb}} \\
Epoch & Mid-transit time\\
&  (BJD - 2,454,900) \\ \hline
\endfirsthead
\caption[]{continued...} \\
Epoch & Mid-transit time \\
& (BJD - 2,454,900) \\ \hline
\endhead
\hline
0& $ 60.9794$ \\
1& $ 74.8155$ \\
2& $ 88.6539$ \\
5& $130.1295$ \\
6& $143.9542$ \\
7& $157.7770$ \\
8& $171.5968$ \\
10& $199.2660$ \\
11& $213.0977$ \\
12& $226.9327$ \\
13& $240.7685$ \\
15& $268.4324$ \\
16& $282.3319$ \\
17& $296.1944$ \\
18& $310.0579$ \\
19& $323.9246$ \\
20& $337.7915$ \\
21& $351.6553$ \\
22& $365.5175$ \\
23& $379.4552$ \\
24& $393.3153$ \\
25& $407.1771$ \\
26& $421.0420$ \\
27& $434.9066$ \\
28& $448.7681$ \\
29& $462.6286$ \\
30& $476.5201$ \\
31& $490.3483$ \\
32& $504.1788$ \\
33& $518.0120$ \\
35& $545.6733$ \\
37& $573.3377$ \\
38& $587.1575$ \\
39& $600.9805$ \\
40& $614.8053$ \\
41& $628.6283$ \\
42& $642.4481$ \\
44& $670.1250$ \\
45& $683.9643$ \\
46& $697.8070$ \\
47& $711.6506$ \\
48& $725.4916$ \\
50& $753.2334$ \\
51& $767.0986$ \\
52& $780.9644$ \\
53& $794.8334$ \\
54& $808.7028$ \\
55& $822.5694$ \\
56& $836.4339$ \\
57& $850.3660$ \\
58& $864.2151$ \\
59& $878.0656$ \\
60& $891.9193$ \\
61& $905.7728$ \\
62& $919.6233$ \\
63& $933.4727$ \\
 \hline
%\end{tabular}

\end{longtable}

\begin{longtable}{lc}
%\centering
%\begin{tabular}{ll}
\caption[]{Planet c mid-transit times from the best-fitting photometric-dynamical model. \label{tab:tabbfc}} \\
Epoch & Mid-transit time \\
& (BJD - 2,454,900) \\ \hline
\endfirsthead
\caption[]{continued...} \\
Epoch & Mid-transit time \\
& (BJD - 2,454,900) \\ \hline
\endhead
\hline
0& $ 55.9117$ \\
1& $ 72.1520$ \\
2& $ 88.3923$ \\
3& $104.6379$ \\
4& $120.8902$ \\
5& $137.1401$ \\
6& $153.3900$ \\
7& $169.6425$ \\
8& $185.8856$ \\
9& $202.1295$ \\
10& $218.3731$ \\
11& $234.6144$ \\
12& $250.8568$ \\
13& $267.1016$ \\
15& $299.5242$ \\
16& $315.7447$ \\
18& $348.1836$ \\
19& $364.4059$ \\
20& $380.5804$ \\
21& $396.8042$ \\
22& $413.0258$ \\
23& $429.2460$ \\
24& $445.4680$ \\
25& $461.6917$ \\
26& $477.8975$ \\
27& $494.1442$ \\
28& $510.3884$ \\
29& $526.6316$ \\
30& $542.8771$ \\
32& $575.3677$ \\
33& $591.6200$ \\
34& $607.8698$ \\
35& $624.1196$ \\
36& $640.3720$ \\
38& $672.8539$ \\
39& $689.0919$ \\
40& $705.3277$ \\
41& $721.5644$ \\
43& $754.0017$ \\
44& $770.2221$ \\
45& $786.4409$ \\
46& $802.6579$ \\
47& $818.8762$ \\
48& $835.0968$ \\
49& $851.2751$ \\
50& $867.5069$ \\
51& $883.7367$ \\
52& $899.9649$ \\
53& $916.1948$ \\
54& $932.4265$ \\
 \hline
%\end{tabular}

\end{longtable}

\begin{figure}
\centering
\includegraphics[width=6in]{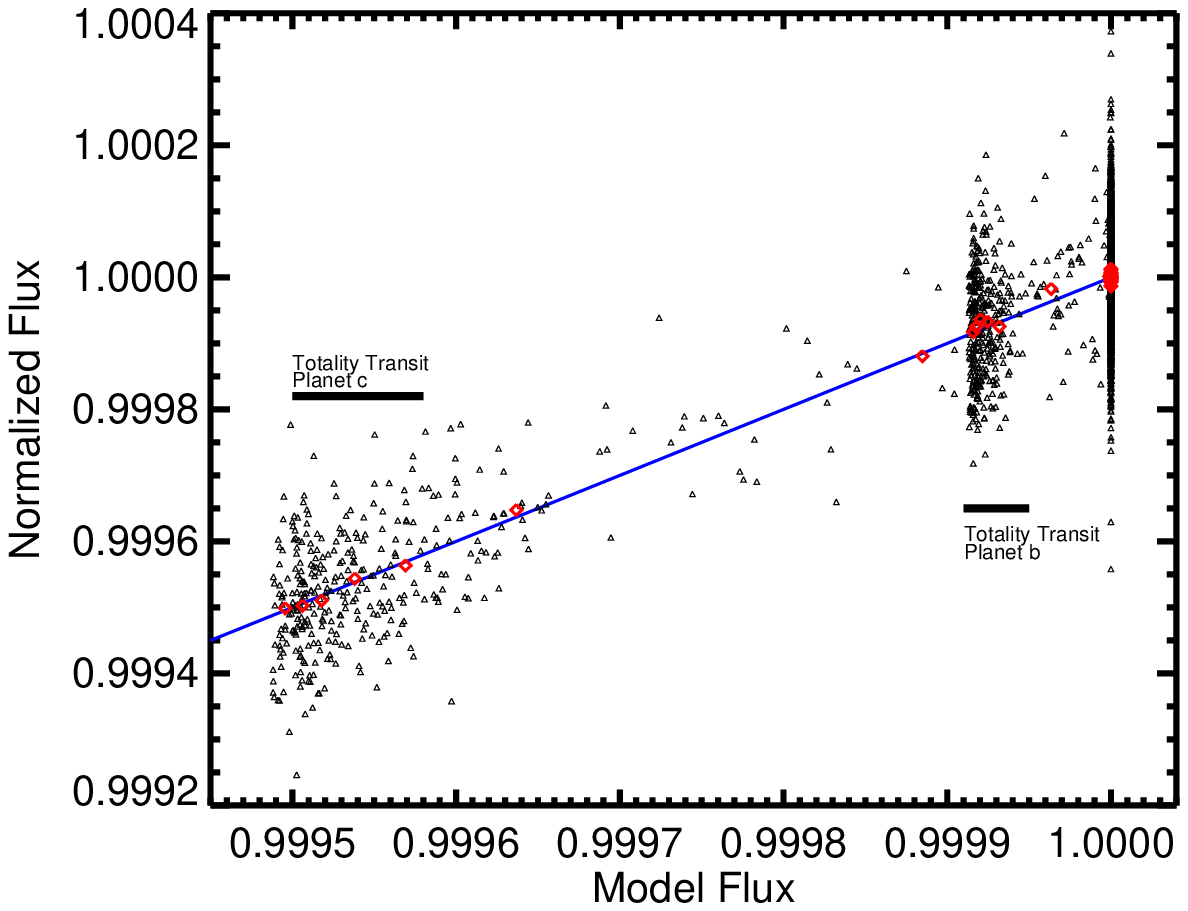}
\caption{Correlation between the photometric-dynamical model flux and normalized
{\it Kepler} photometry (i.e.\ after removal of a local linear fit about the transit).  The
red points are data, sorted in flux, averaged in bins with an equal number of data samples.  The blue line denotes exact correspondence between model and data.  \label{corrflux}}
\end{figure}

Figure \ref{corrflux} shows a plot of the normalized data versus the best-fit
photometric-dynamical model.  The correlation is very nearly linear showing that the
model is an excellent fit to the normalized data. 

\begin{figure}
\centering
\includegraphics[width=6.5in]{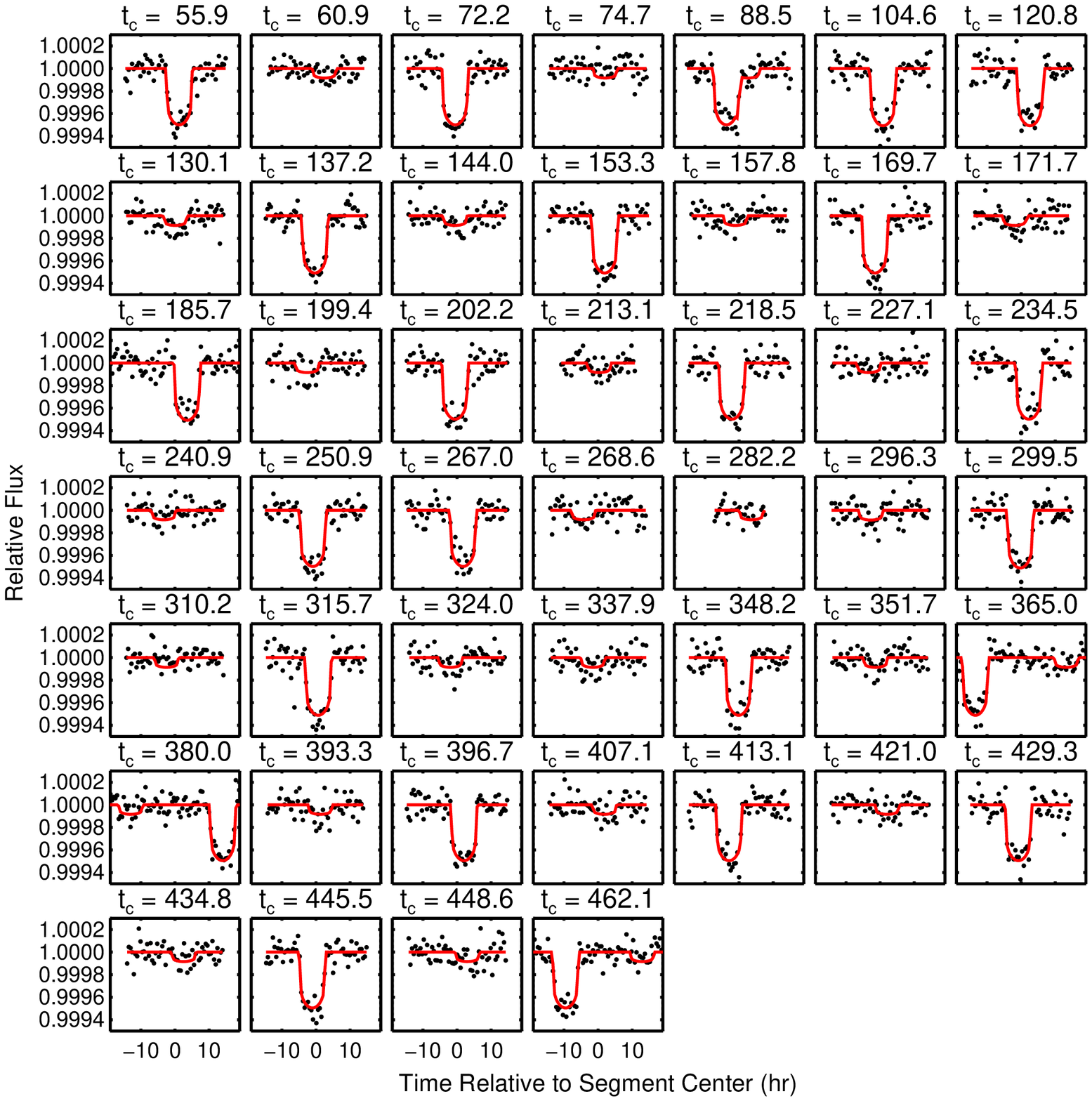}
\caption{Long cadence {\it Kepler} data used in this analysis.  Each panel shows data for a contiguous set of cadences near transits of either Planet b or Planet c.  The time in listed above each panel is near the center time of each panel less 2,454,900 (BJD).  The red curve is the best-fitting photometric-dynamical model.  The time and flux axis is the same in all panels.
  \label{transit_panels_LC}}
\end{figure}

\begin{figure}
\centering
\includegraphics[width=6.5in]{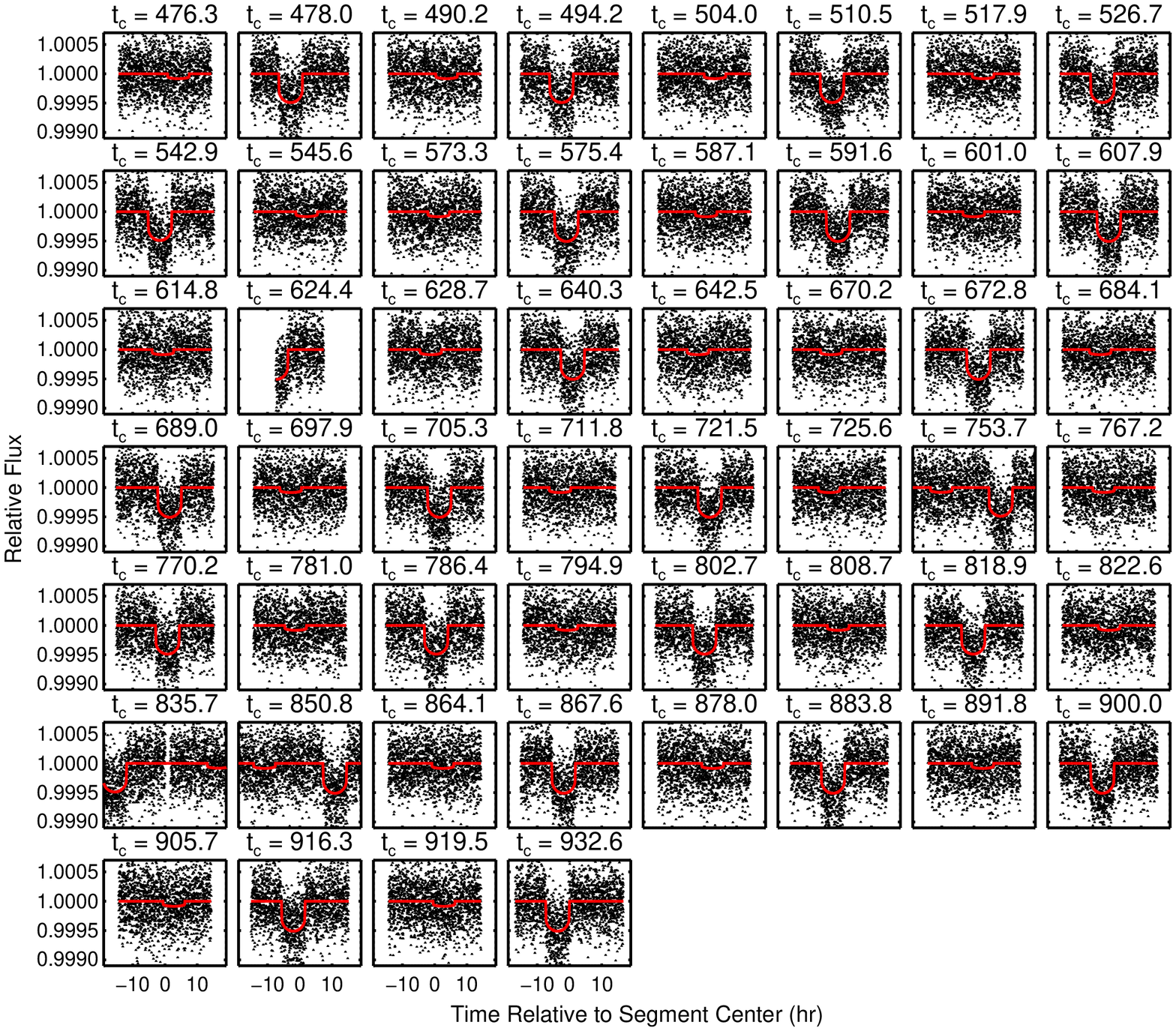}
\caption{Short cadence {\it Kepler} data used in this analysis.  Each panel shows data for a contiguous set of cadences near transits of either Planet b or Planet c.  The time in listed above each panel is near the center time of each panel less 2,454,900 (BJD).  The red curve is the best-fitting photometric-dynamical model.  The time and flux axis is the same in all panels (but larger in flux than Figure \ref{transit_panels_LC} in order to account for the larger relative error in the short cadence data).
  \label{transit_panels_SC}}
\end{figure}

Finally, we have created a set of panels of every transit with the best-fit model in
Figures \ref{transit_panels_LC} and \ref{transit_panels_SC}.  Figure \ref{transit_panels_LC} shows the long cadence data while Figure \ref{transit_panels_SC} shows the short cadence data.  We also have created a movie (Movie S1), attached with this supplement, which shows the
transits of planet c as a function of time. 

The parameters of the best-fit model are listed in Table \ref{tab:tab1} and derived parameters from that best-fit solution are listed in Table \ref{tab:tab2}. In \S \ref{sec:evbest}, we describe the short-term evolution of the best-fit solution.

 \subsubsection{Parameter estimation methodology} \label{sec:mcmc}

We explored the parameter space and estimated the posterior parameter distribution with a Differential Evolution Markov Chain Monte Carlo (DE-MCMC) algorithm \cite{terBraak}. In this algorithm, a large population of independent Markov chains are calculated in parallel.  As in a traditional MCMC, links are added to each chain in the population by proposing parameter jumps, and then accepting or denying a jump from the current state according to the Metropolis-Hastings criterion, using the likelihood function given in \S \ref{sec:like}. What is different from a traditional MCMC is the manner in which jump sizes and directions are chosen for the proposals.  A population member's individual parameter jump vector at step $i+1$ is calculated by selecting two randomly chosen population members (not including itself), and then forming the difference vector between their parameter states at step $i$ and scaling by a factor $\Gamma$.  This is the Differential Evolution component of the algorithm. The factor $\Gamma$ is adjusted such that the fraction of accepted jumps, averaged over the whole population, is approximately 25\%.

We generated a population of 60 chains and evolved through approximately 500,000 generations.  The initial parameter states of the 60 chains were randomly selected from an over-dispersed region in parameter space bounding the final posterior distribution.  The first 10\% of the links in each individual Markov chain were clipped, and the resulting chains were concatenated to form a single Markov chain, after having confirmed that each chain had converged according to the standard criteria including the Gelman-Rubin convergence statistics and the observation of a long effective chain length in each parameter (as determined from the chain autocorrelation).

\subsubsection{Characteristics of the parameter posterior} \label{sec:posterior}

\begin{figure}
\centering
\includegraphics[width=6.5in]{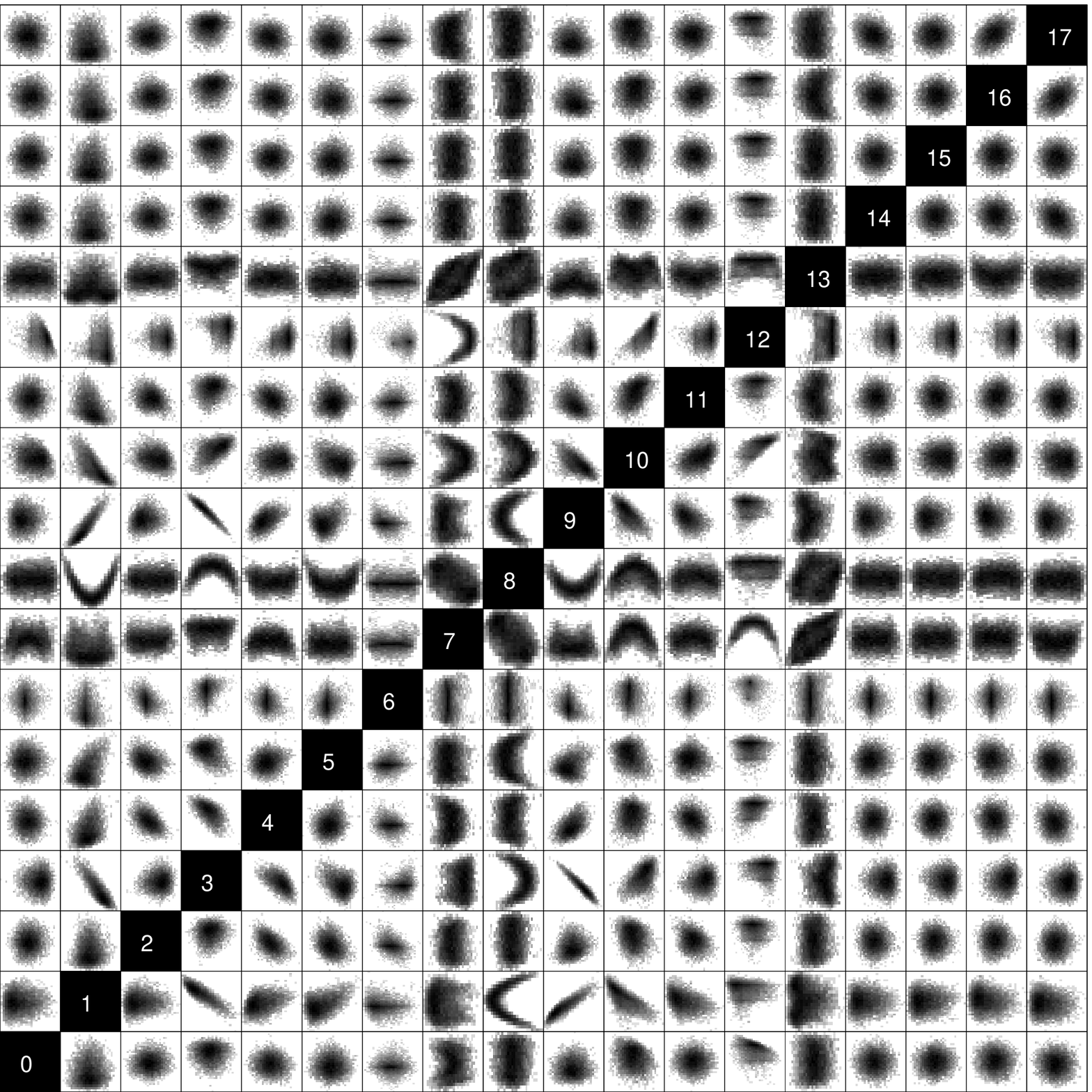}
\caption{Two-parameter joint posterior distributions of the primary model parameters (excluding contamination parameters; see \S \ref{sec:params}).  The densities are plotted logarithmically in order to elucidate the nature of the parameter correlations.  The indices listed along the diagonal indicate which parameter is associated with the corresponding row and column.  The parameter name corresponding to a given index is indicated in Table \ref{tab:tab1} in the ``Index'' column.  \label{corr277}}
\end{figure}

 The parameter values and derived values reported in Tables \ref{tab:tab1}, \ref{tab:tab2}, beside the best-fit values (see \S \ref{sec:best}), were found by computing the 15.8\%, 50\%, 84.2\% levels of the cumulative distribution of the marginalized posterior for each parameter.  Figure \ref{corr277} shows two-parameter joint distributions between all parameters.  This figure is meant to highlight the qualitative features of the posterior as opposed to providing quantitative ranges.  The numbers in that figure correspond to the model parameters in Table \ref{tab:tab1} with the same number listed as ``Index.''  Some non-linear degeneracies still remain in the form of ``bananas" shown in the joint distribution plane, but these appear to have been properly traversed and mixed according to the convergence criteria listed above.
 
 \begin{table}
\centering
\begin{tabular}{|ll|l|lll|}
\hline
Index & Parameter Name & Best-fit & 50\% & 15.8\% & 84.2\% \\ \hline
&~{\it Mass parameters} & & & & \\
0&~~Mean Density, $\rho_\star$ (g cm$^{-3}$) & $0.3531$ & $0.3511 $ & $-0.0052$ & $+0.0053$\\
1&~~Mass sum ratio, $q_+ (\times 10^5)$ & $3.40$ & $3.51 $ & $-0.14$ & $+0.24$\\
2&~~Planetary mass ratio, $q_p$ & $0.553$ & $0.550 $ & $-0.009$ & $+0.010$\\
&~{\it Eccentricity parameters} & & & & \\
4&~~Cosine Difference, $h_-$ & $0.00397$ & $0.00393 $ & $-0.00027$ & $+0.00028$\\
6&~~Cosine Sum, $h_+$ & $0.004$ & $0.004 $ & $-0.018$ & $+0.020$\\
10&~~Sine Difference, $k_-$ & $0.0270$ & $0.0258 $ & $-0.0018$ & $+0.0015$\\
12&~~Sine Sum, $k_+$ & $0.027$ & $0.017 $ & $-0.030$ & $+0.012$\\
&~{\it Planet b Orbit} & & & &\\
9&~~Orbital Period, $P_b$ (day) & $  13.83988$ & $  13.83989 $ & $-0.00060$ & $+0.00082$\\
11&~~Time of Transit, $t_b$ (days since $t_0$) & $10.97981$ & $10.97529 $ & $- 0.00583$ & $+ 0.00548$\\
13&~~Orbital Inclination, $i_b$ (deg) & $  89.52$ & $  90.01 $ & $-0.71$ & $+0.69$\\
&~{\it Planet c Orbit} & & & &\\
3&~~Orbital Period, $P_c$ (day) & $  16.23853$ & $  16.23855 $ & $-0.00054$ & $+0.00038$\\
5&~~Time of Transit, $t_c$ (days since $t_0$) & $ 5.91174$ & $ 5.91315 $ & $- 0.00097$ & $+ 0.00109$\\
7&~~Orbital Inclination, $i_c$ (deg) & $ 89.76$ & $ 89.98 $ & $-0.53$ & $+0.54$\\
8&~~Relative Nodal Longitude, $\Delta \Omega$ (deg) & $  0.25$ & $  0.06 $ & $-1.54$ & $+1.48$\\
&~{\it Radius Parameters} & & & &\\
15&~~Stellar Radius, $R_\star$ ($R_\odot$) & $1.619$ & $1.626 $ & $-0.019$ & $+0.019$\\
16&~~b Radius Ratio, $R_b/R_\star$ & $0.00857$ & $0.00838 $ & $-0.00017$ & $+0.00016$\\
17&~~c Radius Ratio, $R_c/R_\star$ & $0.02079$ & $0.02076 $ & $-0.00018$ & $+0.00018$\\
14&~~Linear Limb Darkening Parameter, $u$ & $0.446$ & $0.456 $ & $-0.024$ & $+0.024$\\
&~{\it Relative Contamination, $F_{\rm cont}/F_\star$} ($\times 100$) & & & &\\
&~~Quarter 1 & $ 1.41$ & $ 1.50 $ & $- 3.10$ & $+ 3.27$\\
&~~Quarter 2 & $-0.20$ & $-0.33 $ & $- 2.46$ & $+ 2.46$\\
&~~Quarter 3 & $ 1.75$ & $ 0.77 $ & $- 2.53$ & $+ 2.61$\\
&~~Quarter 4 & $-0.98$ & $-0.22 $ & $- 2.81$ & $+ 2.79$\\
&~~Quarter 5 & $ 2.34$ & $ 1.97 $ & $- 2.56$ & $+ 2.64$\\
&~~Quarter 6 & $ 3.43$ & $ 2.97 $ & $- 2.32$ & $+ 2.39$\\
&~~Quarter 7 & & & 0 (fixed) &\\
&~~Quarter 8 & $-0.02$ & $-1.41 $ & $- 2.31$ & $+ 2.37$\\
&~~Quarter 9 & $ 3.71$ & $ 3.76 $ & $- 2.27$ & $+ 2.28$\\
&~~Quarter 10 & $-0.72$ & $-2.29 $ & $- 2.02$ & $+ 2.12$\\
&~{\it Noise Parameters} & & & & \\
&~~Long Cadence Relative Width, $\sigma_{\rm LC}$ ($\times 10^5$) & $7.89$ & $7.87 $ & $-0.10$ & $+0.11$\\
&~~Short Cadence Relative Width, $\sigma_{\rm SC}$ ($\times 10^5$) & $  33.692$ & $  33.681 $ & $-   0.077$ & $+   0.078$\\
 \hline
\end{tabular}
\caption{ Model parameters as defined in the text. The reference epoch is $t_0 = $2,454,950 (BJD). \label{tab:tab1}}
\end{table}

\begin{table}
\centering
\begin{tabular}{|l|l|lll|}
\hline
~~Parameter & Best-fit & 50\% & 15.8\% & 84.2\% \\ \hline
~{\it Planetary Bulk Properties} & & & & \\
~~Mass of Planet b, $M_b$ ($M_\oplus$) & $4.28$ & $4.45 $ & $-0.27$ & $+0.33$\\
~~Mass of Planet c, $M_c$ ($M_\oplus$) & $7.73$ & $8.08 $ & $-0.46$ & $+0.60$\\
~~Radius of Planet b, $R_b$ ($R_\oplus$) & $1.512$ & $1.486 $ & $-0.035$ & $+0.034$\\
~~Radius of Planet c, $R_c$ ($R_\oplus$) & $3.669$ & $3.679 $ & $-0.054$ & $+0.054$\\
~~Density of Planet b, $\rho_b$ (g cm$^{-3}$) & $6.80$ & $7.46 $ & $-0.59$ & $+0.74$\\
~~Density of Planet c, $\rho_c$ (g cm$^{-3}$) & $0.860$ & $0.891 $ & $-0.046$ & $+0.066$\\
~~Planetary Density Ratio, $\rho_b/\rho_c$ & $7.91$ & $8.35 $ & $-0.46$ & $+0.51$\\
~~Planet b to Star Density Ratio, $\rho_b/\rho_\star$ & $19.3$ & $21.2 $ & $- 1.6$ & $+ 2.1$\\
~~Planet c to Star Density Ratio, $\rho_c/\rho_\star$ & $2.44$ & $2.53 $ & $-0.12$ & $+0.19$\\
~~Surface Gravity of Planet b, $g_b$ (m s$^{-2}$) & $18.3$ & $19.7 $ & $- 1.3$ & $+ 1.7$\\
~~Surface Gravity of Planet c, $g_c$ (m s$^{-2}$) & $5.62$ & $5.83 $ & $-0.28$ & $+0.42$\\
~~Escape Velocity of Planet b, $v_{{\rm esc},b}$ (km s$^{-1}$) & $18.80$ & $19.34 $ & $- 0.57$ & $+ 0.73$\\
~~Escape Velocity of Planet c, $v_{{\rm esc},c}$ (km s$^{-1}$) & $16.22$ & $16.56 $ & $- 0.42$ & $+ 0.58$\\
~{\it Orbital Properties} & & & & \\
~~Semimajor Axis of Planet b, $a_b$ (AU) & $0.1151$ & $0.1153 $ & $-0.0015$ & $+0.0015$\\
~~Semimajor Axis of Planet c, $a_c$ (AU) & $0.1280$ & $0.1283 $ & $-0.0016$ & $+0.0016$\\
~~$e_b \cos \omega_b+(a_c/a_b) e_c \cos \omega_c$ & $0.004$ & $0.004 $ & $-0.019$ & $+0.021$\\
~~$e_b \cos \omega_b-(a_c/a_b) e_c \cos \omega_c$ & $0.00444$ & $0.00439 $ & $-0.00029$ & $+0.00031$\\
~~$e_b \sin \omega_b+(a_c/a_b) e_c \sin \omega_c$ & $0.027$ & $0.017 $ & $-0.032$ & $+0.013$\\
~~$e_b \sin \omega_b-(a_c/a_b) e_c \sin \omega_c$ & $0.02698$ & $0.02664 $ & $-0.00139$ & $+0.00094$\\
~~Mutual Orbital Inclination, $I$ (deg) & $0.35$ & $1.20 $ & $-0.63$ & $+0.79$\\
~~Orbital Velocity of Planet b, $2\pi a_b/P_b$ (km s$^{-1}$) & $90.4$ & $90.7 $ & $- 1.2$ & $+ 1.2$\\
~~Orbital Velocity of Planet c, $2\pi a_c/P_c$ (km s$^{-1}$) & $85.8$ & $86.0 $ & $- 1.1$ & $+ 1.1$\\
~~Mutual Hill Radius, $R_H$, $\left(\frac{q_+}{24}\right)^{1/3}\left(a_b+a_c\right)$ (AU) & $ 0.002730$ & $ 0.002769 $ & $- 0.000054$ & $+ 0.000067$\\
~{\it Transit Parameters} & & & & \\ 
~~Radius Ratio of Planet b, $R_b/R_\star$ & $0.00857$ & $0.00838 $ & $-0.00017$ & $+0.00016$\\
~~Radius Ratio of Planet c, $R_c/R_\star$ & $0.02079$ & $0.02076 $ & $-0.00018$ & $+0.00018$\\
~~Impact Parameter of Planet b, $b_b/R_\star$ & $ 0.13$ & $-0.00 $ & $- 0.18$ & $+ 0.19$\\
~~Impact Parameter of Planet c, $b_c/R_\star$ & $0.07$ & $0.00 $ & $-0.16$ & $+0.16$\\
~~Transit Velocity of Planet b, $v_b/R_\star$ (day$^{-1}$) & $7.129$ & $7.079 $ & $-0.104$ & $+0.043$\\
~~Transit Velocity of Planet c, $v_c/R_\star$ (day$^{-1}$) & $6.579$ & $6.541 $ & $-0.087$ & $+0.035$\\
~~Transit Duration of Planet b (hr) & $6.74$ & $6.78 $ & $-0.11$ & $+0.09$\\
~~Transit Duration of Planet c (hr) & $7.430$ & $7.443 $ & $-0.020$ & $+0.020$\\
~~Transit Ingress/Egress Duration of Planet b (min) & $3.490$ & $3.455 $ & $-0.086$ & $+0.122$\\
~~Transit Ingress/Egress Duration of Planet c (min) & $ 9.12$ & $ 9.21 $ & $- 0.14$ & $+ 0.27$\\
~~Temperature Scaling of Planet b, $\sqrt{R_\star/2a_b}$ & $0.18087$ & $0.18104 $ & $-0.00045$ & $+0.00045$\\
~~Temperature Scaling of Planet c, $\sqrt{R_\star/2a_c}$ & $0.17149$ & $0.17165 $ & $-0.00043$ & $+0.00043$\\
 \hline
\end{tabular}
\caption{ Derived Parameters. The reference epoch is $t_0 = $2,454,950 (BJD).\label{tab:tab2}}
\end{table}
 
\begin{figure}
\centering
\includegraphics[width=5in]{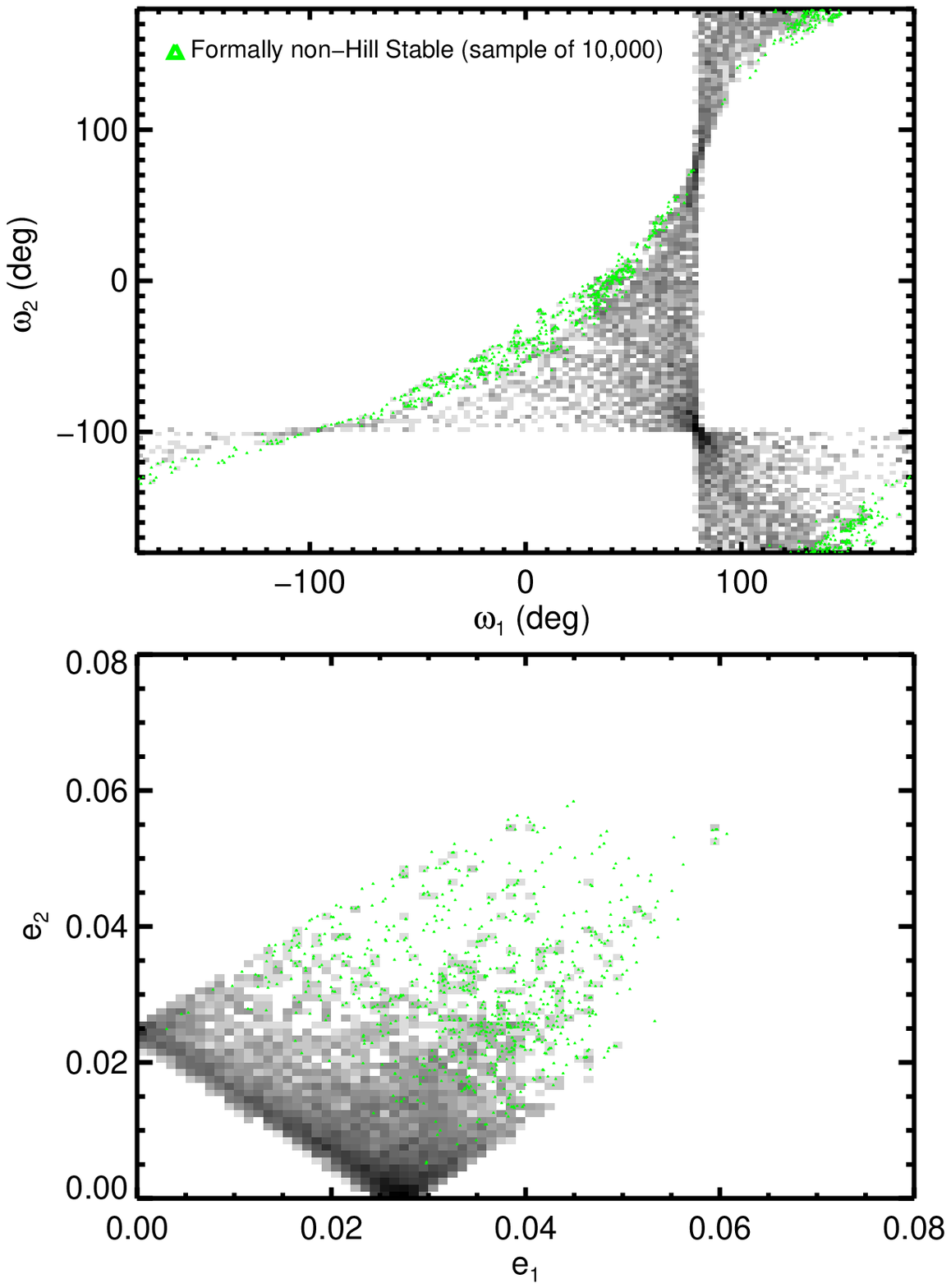}
\caption{Posterior distributions in the eccentricity and argument of pericenter planes.\label{fig:eccomega} }
\end{figure}

Figure \ref{fig:eccomega} shows the posterior distribution in the eccentricity and argument of pericenter planes (this figure is discussed further in \S \ref{sec:stab}).  A number of arguments are permitted, however, only small eccentricities are allowed.  At 95\% confidence, $e_b < 0.039$ and $e_c < 0.033$.  
 
\begin{figure}
\includegraphics[width=6.5in]{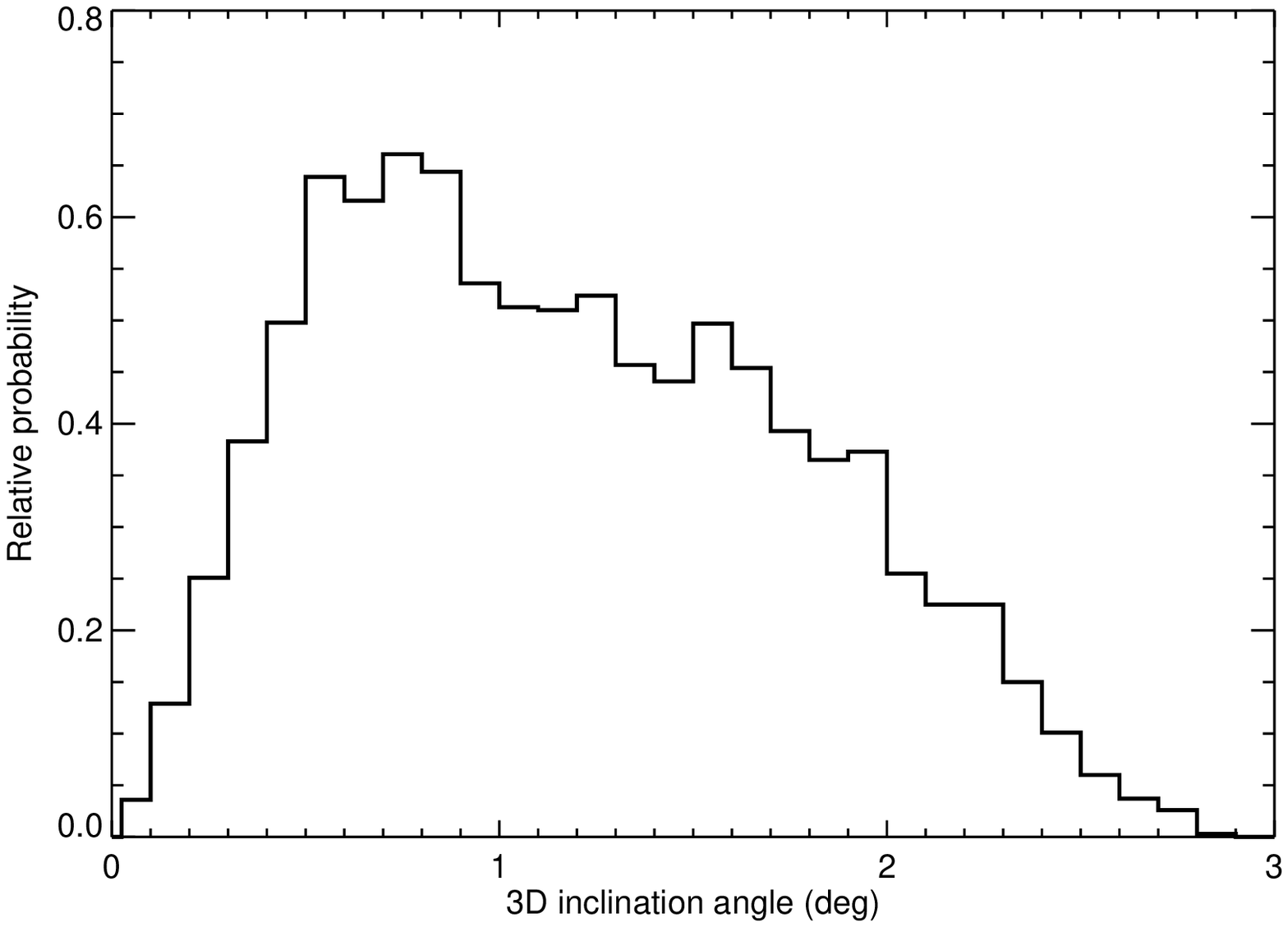}
\caption{Posterior distribution of the mutual inclination, $I$, of the
planetary orbits. \label{inclination_angle}}
\end{figure}

\begin{figure}
\includegraphics[width=6.5in]{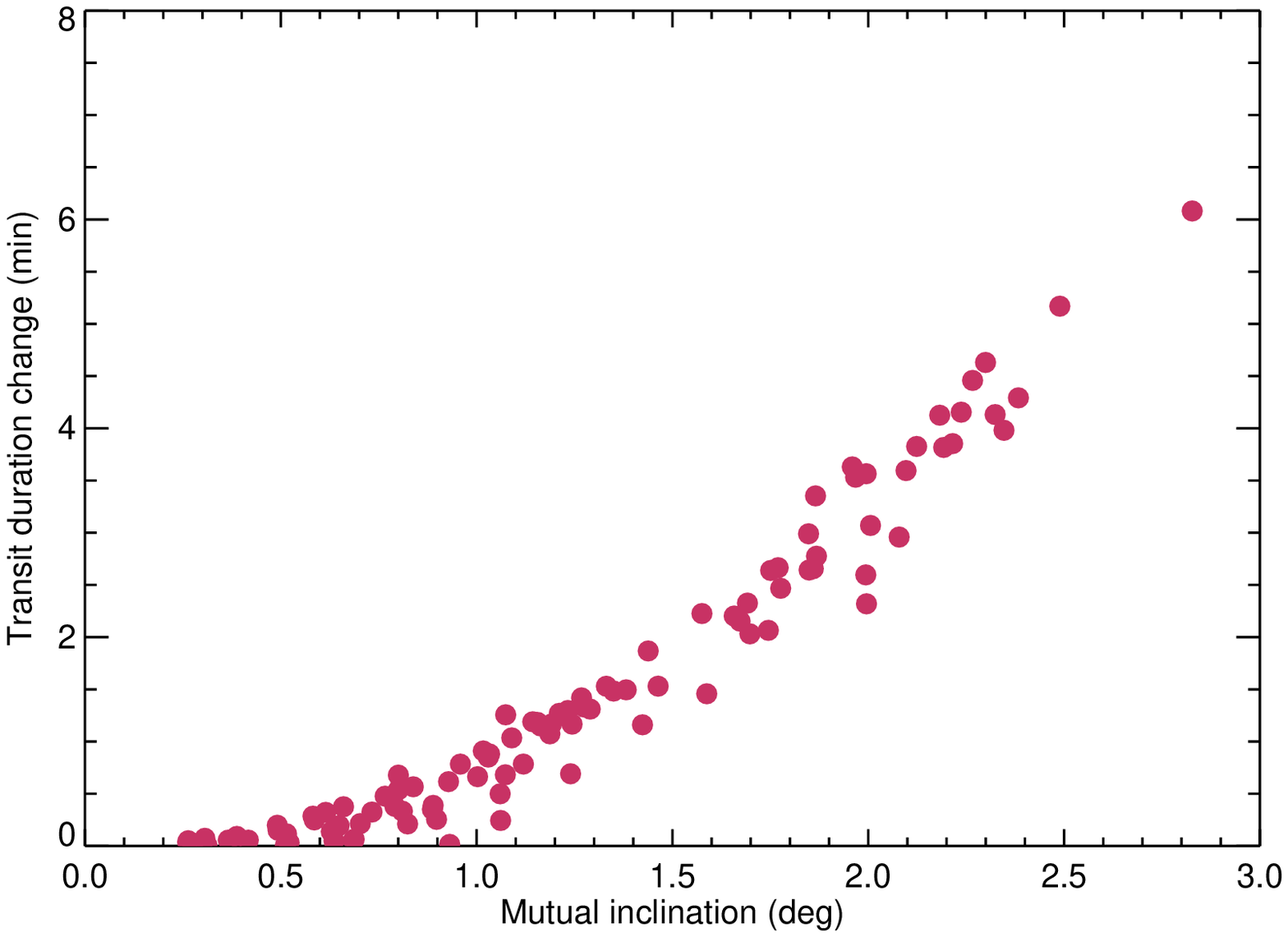}
\caption{\addition{The change in the transit duration (secular component) over the span of observations as a function of mutual inclination for a random sample of 100 initial conditions from the posterior. } \label{fig:tdv}}
\end{figure}

The three-dimensional ``mutual'' inclination between the planets' orbits, which may be defined as an osculating term at the reference epoch as
\begin{eqnarray}
	\cos I & = & \sin i_b \sin i_c \cos \Delta \Omega + \cos i_c \cos i_b,
\end{eqnarray}
 is constrained to within a couple of degrees.  Figure \ref{inclination_angle}
shows the distribution of $I$; Table \ref{tab:tab2} gives confidence intervals in the mutual inclination.  \addition{The constraint on the mutual inclination derives from the lack of significant transit durations variations in the transits of planet c.  To show this, we examined the long term variation of the sky-plane inclinations of each planet based on a draw of 100 initial conditions from our posterior (short term evolution of the posterior is described further in \S~\ref{sec:short}).  The inclinations show circulation of a forced inclination about a free inclination.  We isolated and fitted the secular component (over the time scale of our observations) so as to compute the transit duration variation due to the secular change for planet c as a function of mutual inclination.  We plot this for our 100 initial conditions in Figure~\ref{fig:tdv}.  This figure shows a clear correlation between transit duration change and mutual inclination.  For a mutual inclination of $\approx$2.5 degrees - at the tail of our marginalized posterior distribution in that parameter (see Figure~\ref{inclination_angle}) - the transit duration changes by about 5 minutes over the roughly 900 days of observation or  approximately 0.1 minutes/epoch.   We compared this rate with what can be constrained from the data:  A simple transit model fit to individual transit epochs of planet c shows that the measured uncertainty in duration is approximately 10 minutes at each epoch.  Fitting a linear model to the durations of 50 transits of c (the number that was observed), assuming zero transit duration variation, would yield a 1-sigma uncertainty in the linear model slope of 0.1 minutes/epoch.  The near equivalence between these two numbers - the 1-sigma constraint from the observed durations with assumed-flat transit duration variation and the numerically estimated rate assuming the 1-sigma value of mutual inclinations - demonstrates our claim.  }

 \section{Bulk composition of the Kepler-36 planets} \label{sec:planets}

\subsection{Bayesian Constraints on Kepler-36b's Bulk Composition}

In this section we elaborate briefly on the analysis constraining Kepler-36b's composition.

Following the Bayesian approach outlined in \cite{Rogers7122010}, we account for the (correlated) observational uncertainties in the planet mass and radius in a rigorous way when constraining the planet composition. We adopt a uniform prior on the planet mass and a flat prior on planet composition. The likelihood function (which quantifies how well a specific combination of planet mass and composition fit the observed data) is set by the joint $M_p$--$R_p$ posterior derived from MCMC fits to the transit timing variations and transit light curves (illustrated in Figure 3 in the main text). We calculate planet radii at a given mass and composition using the interior structure model of \cite{Rogers7162010, Rogers7382011}. 

We first explore a rocky planet scenario for Kepler-36b, in which the planet consists of a metal iron core surrounded by a Mg$_{0.9}$Fe$_{0.1}$SiO$_3$ silicate mantle. Our main conclusion is that Kepler-36b could be a rocky planet with an Earth-like composition (with $\sim30$\% of its mass in an iron core, and $\sim70$\% of its mass in a silicate mantle).  Marginalizing over planet mass, the iron core mass fraction of Kepler-36b is constrained. We find median and  68\% credible intervals of $M_{\rm core}/M_p = 0.29^{+0.11}_{- 0.10}$.

We turn to the possibility that Kepler-36b harbors a water envelope that contributes to its transit radius. We follow a similar Bayesian analysis approach as described above, but now consider a three component model for the planet structure (iron core, silicate mantle, water envelope) with a flat prior on the fraction of the planet mass in each layer.
 In this three component composition model, inherent degeneracies come to bare on the composition constraints; many compositions agree with a single planet mass and radius.
The mass of a water envelope on Kepler-36b is less than 23\% of the total planet mass, at 68\% Bayesian confidence. The most water rich composition within the 68\% credible region that have iron abundances below the \cite{Marcus7122010} constraints from silicate collisional stripping simulations is $13\%$ H$_2$O.   

Finally, considering the possibility for a H/He gas layer on Kepler-36b, we find that Kepler-36b could have no more than 1\% of its mass in solar composition H/He layer. 

\subsection{Bayesian constraints on Kepler-36c's composition}

Kepler-36c has a low enough mean density that it must have a hydrogen-rich envelope (a water envelope is not puffy enough). In the main text (and elaborated here in the supplementary online material) we have presented evolution calculations to constrain the H/He envelope mass. Assuming a hot-start formation process and passive cooling at the planet's current orbital separation, tight constraints on Kepler-36c's interior entropy and H/He mass obtain. The formation mechanism and the full dynamical and insolation histories of the Kepler-36 planets are not known, however.

\begin{figure}
\includegraphics[width=6.5in]{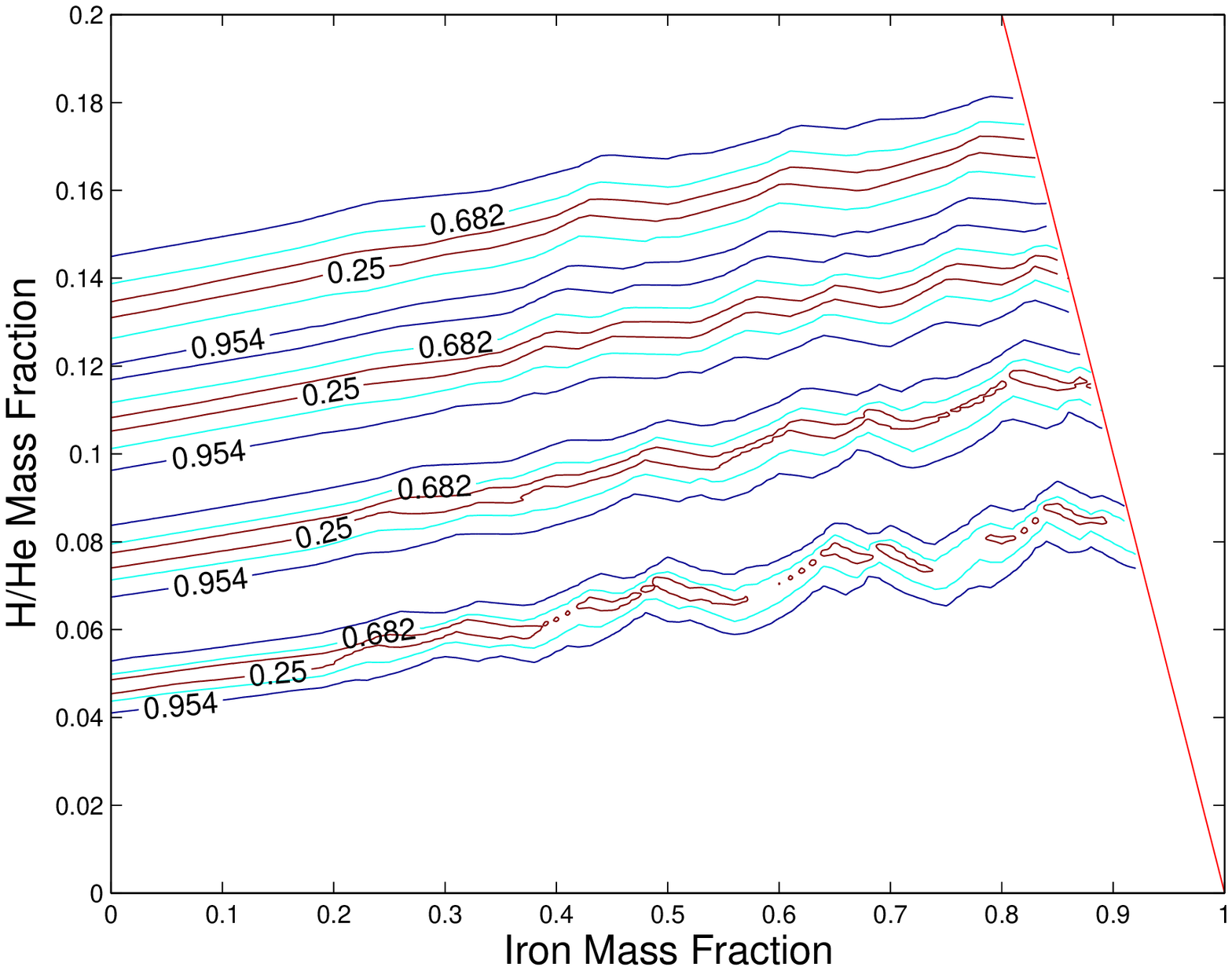}
\caption{Bayesian constraints on the mass of a solar composition envelope surrounding Kepler-36c assuming a rocky interior. The contours plotted are lines of constant probability labeled with a Bayesian confidence value indicating the Òdegree of beliefÓ (between 0 and 1) that the true composition of the planet falls within the contour (given the assumed priors). A rocky heavy element interior comprised of an metal iron core and silicate mantle is assumed. The Bayesian contours in Figures~\ref{fig:K42cRock} and ~\ref{fig:K42cRockIce} are not marginalized over the planet evolution history, but instead show composition constraints for four different choices of Kepler-36c's intrinsic luminosity. From high to low H/He content (low to high $L_p$) the values shown are: $L_p/M_p = 10^{-13}~\rm{W\,kg^{-1}}$, a very low intrinsic luminosity to set strong upper limits on the mass of H/He; $L_p/M_p = 10^{-11.5}~\rm{W\,kg^{-1}} \approx~L_{\rm radio}/M_p$, the expected luminosity from the decay of radiogenic isotopes at 6.8~Gyr assuming bulk Earth abundances;  $L_p/M_p = 10^{-10.5}~\rm{W\,kg^{-1}} \approx~10L_{\rm radio}/M_p$; and $L_p/M_p = 10^{-9.5}~\rm{W\,kg^{-1}} \approx~100L_{\rm radio}/M_p$.}
\label{fig:K42cRock}
\end{figure}

\begin{figure}
\includegraphics[width=6.5in]{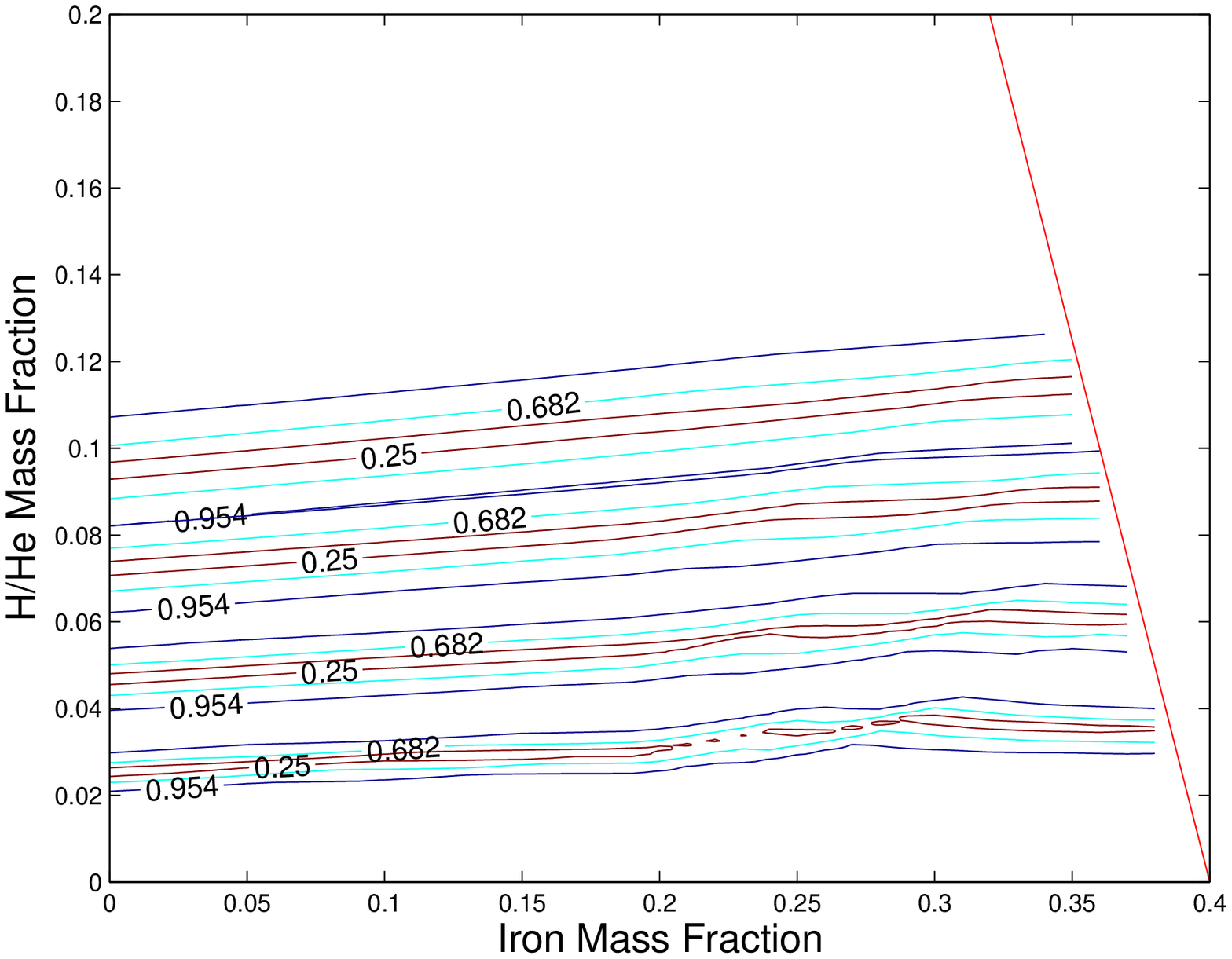}
\caption{Bayesian constraints on the mass of a solar composition envelope surrounding Kepler-36c assuming an ice-rock interior. This figure is identical to Figure~\ref{fig:K42cRock}, but assumes the  heavy element interior of Kepler-36c is 60\% H$_2$O and 40\% rock by mass. The iron to silicate ratio abundance is still allowed to vary.}
\label{fig:K42cRockIce}
\end{figure}

In this section we complement the planet cooling calculations with a more expansive investigation of parameter space, to illustrate the dependence of the planet compositional inference on the intrinsic luminosity of the planet. In cases where the planet history is more complicated than the isolated cooling calculations have assumed, a systematic offset in the prediction of $L_p(t)$ could result. The Bayesian contours in Figures~\ref{fig:K42cRock} and ~\ref{fig:K42cRockIce} offer Kepler-36c composition constraints as a function of the planet intrinsic luminosity.

We consider two scenarios for the composition of the planet's heavy-element interior: a rocky (iron and silicate) interior (Figure~\ref{fig:K42cRock}), and an ice-rock interior consisting of 60\% H$_2$O and 40\% rock by mass (Figure~\ref{fig:K42cRockIce}). In both scenarios we consider a range of iron to silicate mass ratios, and assume a uniform prior on the mass fraction within each compositional layer in the planet. Four different choices of Kepler-36c's intrinsic luminosity are shown for illustration: $L_p/M_p = 10^{-13}~\rm{W\,kg^{-1}}$, a very low intrinsic luminosity to set strong upper limits on the mass of H/He; $L_p/M_p = 10^{-11.5}~\rm{W\,kg^{-1}} \approx~L_{\rm radio}/M_p$, the expected luminosity from the decay of radiogenic isotopes at 6.8~Gyr assuming bulk Earth abundances;  $L_p/M_p = 10^{-10.5}~\rm{W\,kg^{-1}} \approx~10L_{\rm radio}/M_p$; and $L_p/M_p = 10^{-9.5}~\rm{W\,kg^{-1}} \approx~100L_{\rm radio}/M_p$. For low-mass planets with gas layers, the intrinsic luminosity of the planet can have an important effect on the ``puffiness" of the gas layer. The H/He envelope mass fraction inferred for Kepler-36c is smaller for higher entropy, high $L_p$ cases. The hot-start evolution calculations favor the higher luminosity scenarios.

The Bayesian composition constraint contours in Figures~\ref{fig:K42cRock} and ~\ref{fig:K42cRockIce} provide a lower bound on the uncertainties in Kepler-36c's interior composition and H/He envelope mass at a single snapshot of planet intrinsic luminosity. Marginalizing over the planet cooling history to account for the uncertainty in $L_{p}\left(t\right)$ will further broaden these composition constraints. Uncertainties quoted in the cooling calculation results are not credible intervals like those presented in Figures~\ref{fig:K42cRockIce} and \ref{fig:K42cRock}, but are instead estimated by varying one parameter ($M_p$, $R_p$, $C_p$, age, incident flux) at a time and adding in quadrature. The heavier computational burden of the evolution calculation currently precludes a full Bayesian analysis of the composition uncertainties.
Ultimately, future work will incorporate cooling calculations in the Bayesian approach to determine priors on $L_{p}\left(t\right)$ and the planet interior entropy.

\subsection{Constraints on Kepler-36c considering thermal evolution}
Kepler-36c is low density, substantially less dense than pure water. As a result, it must have a substantial Hydrogen \& Helium (H/He) envelope. In order to estimate the planet's mass fraction in H/He, we use a coupled thermal evolution and mass loss code \cite{Lopez2012}. These models are based on the model used in \cite{Miller7362011}, which has been adapted for modeling Super-Earths following the methods of \cite{Nettelmann7332011}. By fully modeling the thermal evolution of the planet's interior, we are able to get a much more precise estimate of the the planet's interior structure. Models that only compute an instantaneous structure are forced to vary the intrinsic luminosity of the planet over several orders of magnitude, which introduces large errors in the composition.

In particular, the model assumes an adiabatic H/He interior structure that initially has a large entropy from formation. We then track the planet's radius up to the present day as it cools and contracts. In general, the planet's radius is insensitive to the initial entropy choice by 100 Myr. The bottom of the H/He adiabat then attaches to an isothermal core with a Earth-like rock/iron ratio. Meanwhile, the radiative top of the atmosphere also becomes isothermal once this intrinsic flux from this adiabat equals the incident flux at the top of the atmosphere. We find cooling rates at the top of the atmosphere, by interpolating between models of the intrinsic flux computed on a grid of gravity, Teff, and entropy. These cooling rates were computed using the self-consistent, non-gray radiative transfer models described in \cite{Fortney6592007}, assuming cloud-free 50x solar opacities.

For the rocky core we use the ANEOS olivine equation of state (EOS) \cite{Thompson1990} for the rock and SESAME FeEOS \cite{Lyon1992} for the iron. For the H/He envelope we use SCVH \cite{Saumon991995} and for $H_2O$ we use H2O-REOS \cite{Nettelmann6832008}.

Finally, we model the cooling of the rocky core. As the H/He adiabat cools and contracts, the isothermal core needs to cool as well. However, because this rock/iron core has a nonzero heat capacity, the presence of a large core will slow down the cooling of the atmosphere. Although this effect is negligible for gas giants, it becomes dominant for Super-Earths. Models that neglect the cooling of the core will predict interiors that are too cold and over predict amount of volatiles needed to match the observed radius. We vary the heat capacity of the rock from $0.5-1.0$ J/K/g as in \cite{Nettelmann7332011}. We also included radiogenic heating in the rocky core, assuming earth like abundances of U, Th, and K.

In order to determine the current composition of Kepler-36c, we ran thermal evolution models without mass loss and adjusted the mass of the core until we matched the observed radius at the current age. We then determined the uncertainty in the current composition due to the observed uncertainties in the planet's mass, radius, age, and incident flux along with the theoretical uncertainties in the atmospheric albedo, the heat capacity and the rock/iron ratio of the core. We varied the albedo from $0-0.8$ and the iron fraction from pure rock to the maximum possible iron fraction from collisional stripping \cite{Marcus7002009}. The dominant sources of error were the planet radius, the iron fraction, and the heat capacity of the core, all others were negligible. Finally, we were able to get a estimate of the current composition of Kepler-36c of $8.6\pm^{1.4}_{1.3}\%$  H/He, assuming a water-free interior.

We also explored the possibility of a water-rich interior for Kepler-36c. To do this we inserted a water layer equal in mass to the rocky core in between the rock/iron core and the H/He envelope. From the thermal evolution, we find that this water layer should be in the molecular and ionic fluid phases and not in high pressure ices. We find that for a water-rich, or ``Neptune-like'' composition, Kepler-36c needs $1.6\pm0.4\%$ of its mass in H/He.

Despite the large density contrast seen today, it is possible both planets could have been volatile rich in the past. Due their high incident fluxes, both planets are vulnerable to XUV driven atmospheric escape of light gases \cite{Lammer5982003, Erkaev4722007}. Although it is likely rocky today, we find that Kepler-36b could have been as much as $40\%$ H/He when the system was 100 Myr old and the stellar XUV flux was $\sim 200$ times higher \cite{Ribas6222005}. With similar models, we find that Kepler-36c would have been $35\pm9\%$ H/He, assuming a dry interior.

\section{Short-term behavior and secular characteristics \\of Kepler-36} \label{sec:short}

\subsection{Evolution of the best-fit model} \label{sec:evbest}

\begin{figure}
\includegraphics[width=6.5in]{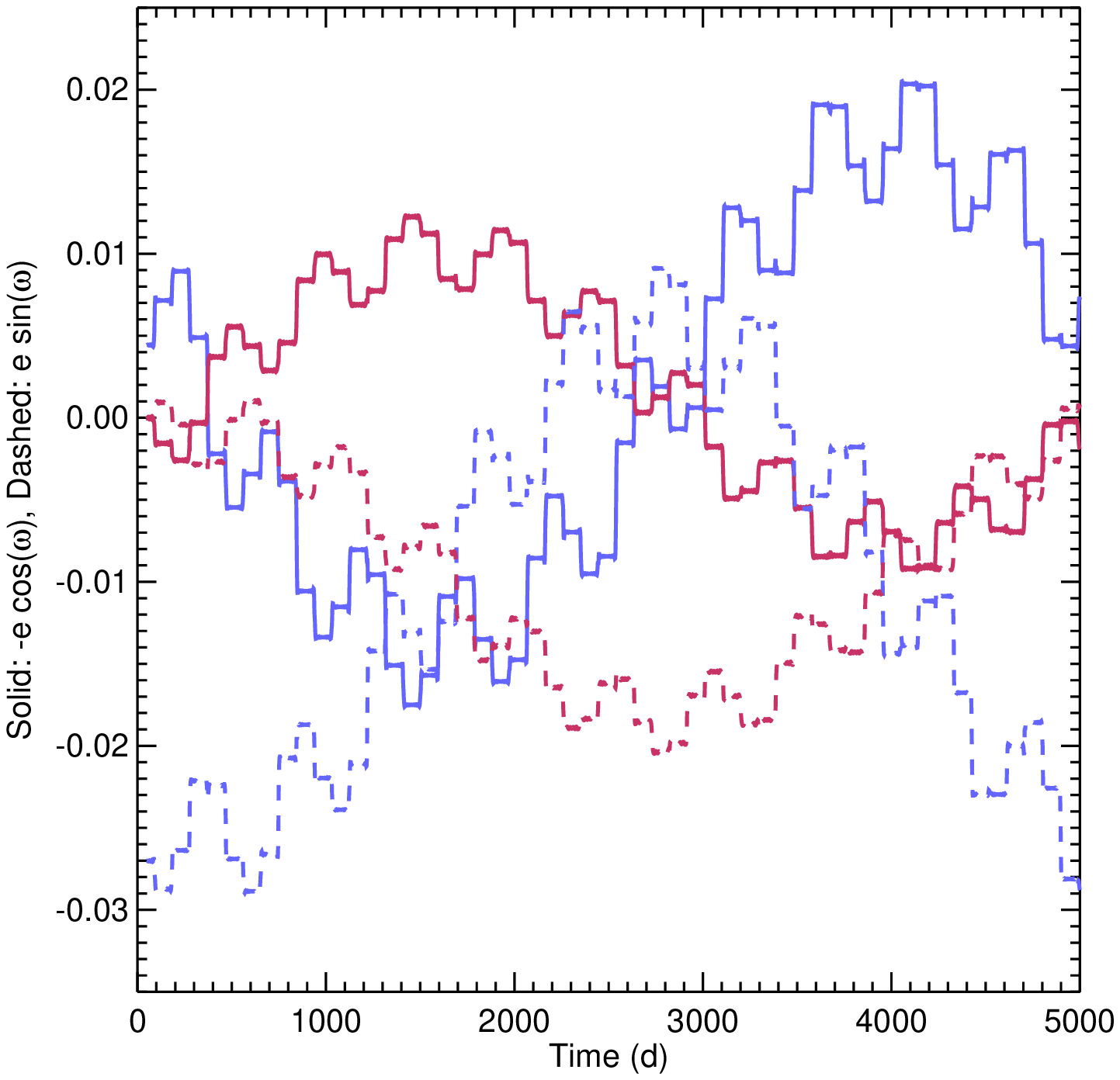}
\caption{Plot of eccentricity vectors for the two planets as a function
of time.  Kepler-36b is plotted in blue, while Kepler-36c is plotted in
red.  \label{e_omega}}
\end{figure}

\begin{figure}
\includegraphics[width=6.5in]{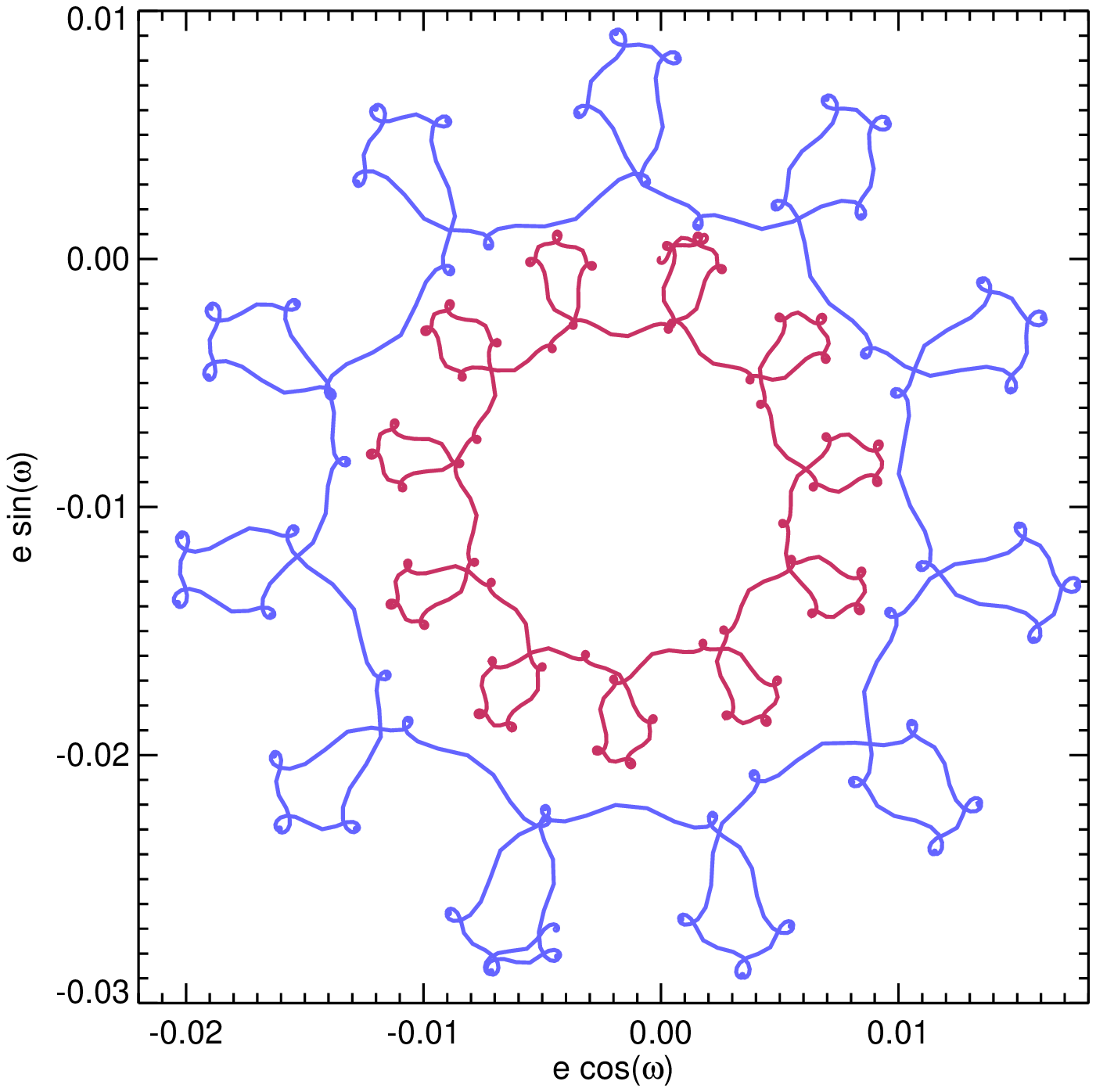}
\caption{Plot of the two components of the eccentricity vector versus
one another.  Kepler-36b is plotted in blue, while Kepler-36c is plotted in
red.  \label{evector}}
\end{figure}

The orbital elements show oscillatory patterns as a function of
time.  Figure \ref{e_omega} shows the eccentricity vectors plotted
versus time. The expected variation in eccentricity is $
 \approx 0.01$, which is indeed the case for the
best-fit model.  The eccentricity variation of the smaller planet
is larger by a factor of the mass ratio, $\approx 1.8$, which is
expected due to conservation of momentum during the conjunctions.
The orbital elements stay approximately constant
in between conjunctions, then show a jump as they pass through
conjunction.   As the orientations of the conjunctions circulate
with time, the orbital elements oscillate.  When the components
of the eccentricity vectors are plotted versus one another,
Figure \ref{evector}, the
pattern becomes rather intricate.  The orbital elements remain
nearly constant between conjunctions creating the knots in
each petal.  During conjunction the planets undergo the largest
change in orbital elements, forming the wiggles connecting the
knots.  The conjunctions drift in inertial space over $\approx
5$ conjunctions;  however, the planets have slightly different
orbital elements after 5 conjunctions, so the petals drift
and eventually form the entire ``flower."

\begin{figure}
\includegraphics[width=6.5in]{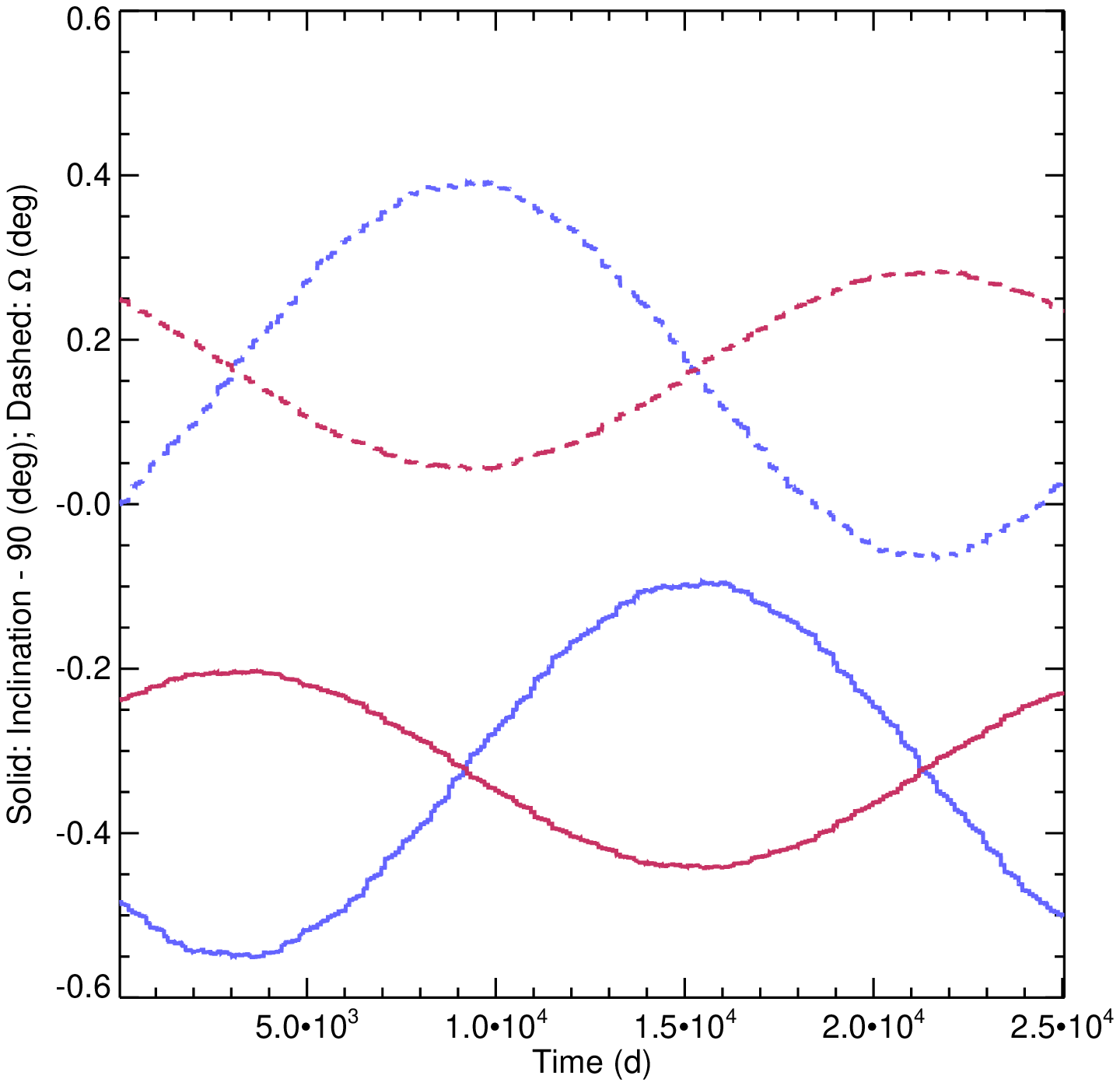}
\caption{Plot of inclination angles for the two planets and nodal angles as a function
of time.  Kepler-36b is plotted in blue, while Kepler-36c is plotted in
red.  \label{inclination_capomega}}
\end{figure}

The planet orbits precess with time;  Figure \ref{inclination_capomega}
shows the inclination angle and sky nodal angle for one of the
best-fit models.  The timescale of precession in this case is
$\approx 2.5 \times 10^4$ days.

 \subsection{Predicted Ephemerides and transit parameters}

Tables \ref{tab:tab3} and \ref{tab:tab4} provide predicted times of transit, impact parameters and duration for the next year (starting near May 1, 2012) with errors.   Figure \ref{tdv_predicted} plots predictions over a much longer time range after subtracting a linear model.  Figure \ref{tdv_predicted} shows the variations in the transit durations during that same time.

\begin{table}
\centering
{
\begin{tabular}{|l|l|l|l|}
\hline
Epoch, $E$ & $T_{\rm transit}-(\langle T_0 \rangle+E\langle P\rangle )$ (hr) & Impact Parameter & Duration (hr) \\ \hline
0 & $ 0.98\pm 0.22$ & $0.130\pm0.078$ & $6.783\pm$0.077\\
1 & $ 0.10\pm 0.22$ & $0.130\pm0.078$ & $6.783\pm$0.077\\
2 & $-0.32\pm 0.22$ & $0.130\pm0.078$ & $6.809\pm$0.077\\
3 & $-0.47\pm 0.21$ & $0.130\pm0.078$ & $6.807\pm$0.077\\
4 & $-0.67\pm 0.21$ & $0.130\pm0.078$ & $6.808\pm$0.077\\
5 & $-0.78\pm 0.21$ & $0.130\pm0.078$ & $6.808\pm$0.077\\
6 & $-0.87\pm 0.21$ & $0.130\pm0.078$ & $6.808\pm$0.077\\
7 & $-1.01\pm 0.21$ & $0.130\pm0.078$ & $6.808\pm$0.077\\
8 & $-1.23\pm 0.21$ & $0.130\pm0.078$ & $6.807\pm$0.077\\
9 & $-0.10\pm 0.19$ & $0.129\pm0.079$ & $6.856\pm$0.078\\
10 & $ 0.02\pm 0.19$ & $0.129\pm0.079$ & $6.855\pm$0.078\\
11 & $ 0.15\pm 0.19$ & $0.129\pm0.079$ & $6.856\pm$0.078\\
12 & $ 0.37\pm 0.19$ & $0.129\pm0.079$ & $6.856\pm$0.078\\
13 & $ 0.59\pm 0.20$ & $0.129\pm0.079$ & $6.856\pm$0.078\\
14 & $ 0.75\pm 0.21$ & $0.129\pm0.079$ & $6.855\pm$0.078\\
15 & $ 0.85\pm 0.22$ & $0.129\pm0.079$ & $6.855\pm$0.078\\
16 & $ 2.31\pm 0.23$ & $0.129\pm0.081$ & $6.872\pm$0.079\\
17 & $ 1.78\pm 0.25$ & $0.129\pm0.081$ & $6.872\pm$0.079\\
18 & $ 1.28\pm 0.26$ & $0.129\pm0.081$ & $6.873\pm$0.079\\
19 & $ 0.86\pm 0.28$ & $0.129\pm0.081$ & $6.873\pm$0.079\\
20 & $ 0.44\pm 0.30$ & $0.129\pm0.081$ & $6.873\pm$0.079\\
21 & $-0.06\pm 0.32$ & $0.129\pm0.081$ & $6.872\pm$0.079\\
22 & $-0.59\pm 0.34$ & $0.129\pm0.081$ & $6.873\pm$0.079\\
23 & $-0.51\pm 0.37$ & $0.128\pm0.081$ & $6.853\pm$0.078\\
24 & $-1.48\pm 0.38$ & $0.128\pm0.081$ & $6.853\pm$0.078\\
25 & $-2.38\pm 0.39$ & $0.128\pm0.081$ & $6.853\pm$0.078\\
 \hline
\end{tabular}
	}
\caption{Predicted ephemerides, impact parameters and durations for Kepler-36b for one year.  In the table, times are given relative to a linear ephemeris with $\langle T_0\rangle=2456049.39050$ (BJD) and $\langle P \rangle = 13.85940$ days.  Epoch $E=0$ is near May 1, 2011.  \label{tab:tab3} }
\end{table}

\begin{table}
\centering
{
\begin{tabular}{|l|l|l|l|}
\hline
Epoch, $E$ & $T_{\rm transit}-(\langle T_0 \rangle+E\langle P\rangle )$ (hr) & Impact Parameter & Duration (hr) \\ \hline
0 & $-0.81\pm 0.05$ & $0.096\pm0.062$ & $7.438\pm$0.019\\
1 & $-0.18\pm 0.05$ & $0.096\pm0.062$ & $7.438\pm$0.019\\
2 & $ 0.19\pm 0.04$ & $0.095\pm0.062$ & $7.423\pm$0.020\\
3 & $ 0.33\pm 0.04$ & $0.095\pm0.062$ & $7.424\pm$0.021\\
4 & $ 0.46\pm 0.03$ & $0.095\pm0.062$ & $7.423\pm$0.021\\
5 & $ 0.53\pm 0.03$ & $0.095\pm0.062$ & $7.423\pm$0.021\\
6 & $ 0.62\pm 0.03$ & $0.095\pm0.062$ & $7.424\pm$0.021\\
7 & $ 0.77\pm 0.03$ & $0.095\pm0.062$ & $7.424\pm$0.021\\
8 & $ 0.11\pm 0.03$ & $0.093\pm0.063$ & $7.395\pm$0.024\\
9 & $ 0.03\pm 0.04$ & $0.093\pm0.063$ & $7.395\pm$0.024\\
10 & $-0.07\pm 0.05$ & $0.093\pm0.063$ & $7.394\pm$0.024\\
11 & $-0.23\pm 0.06$ & $0.093\pm0.063$ & $7.394\pm$0.024\\
12 & $-0.36\pm 0.07$ & $0.093\pm0.063$ & $7.395\pm$0.024\\
13 & $-0.42\pm 0.09$ & $0.093\pm0.063$ & $7.395\pm$0.024\\
14 & $-1.32\pm 0.10$ & $0.092\pm0.064$ & $7.384\pm$0.027\\
15 & $-0.93\pm 0.11$ & $0.092\pm0.064$ & $7.384\pm$0.027\\
16 & $-0.59\pm 0.12$ & $0.092\pm0.064$ & $7.384\pm$0.027\\
17 & $-0.28\pm 0.14$ & $0.092\pm0.064$ & $7.384\pm$0.027\\
18 & $ 0.06\pm 0.15$ & $0.092\pm0.064$ & $7.384\pm$0.027\\
19 & $ 0.45\pm 0.17$ & $0.092\pm0.064$ & $7.384\pm$0.027\\
20 & $ 0.47\pm 0.18$ & $0.091\pm0.064$ & $7.396\pm$0.028\\
21 & $ 1.17\pm 0.19$ & $0.091\pm0.064$ & $7.396\pm$0.028\\
 \hline
\end{tabular}
	}
\caption{Predicted ephemerides, impact parameters and durations for Kepler-36c for one year.  In the table, times are given relative to a linear ephemeris with $\langle T_0\rangle=2456044.91461$ (BJD) and $\langle P \rangle = 16.22407$ days.  Epoch $E=0$ is near April 27, 2011.  \label{tab:tab4} }
\end{table}

\begin{figure}
\centering
\includegraphics[width=6.5in]{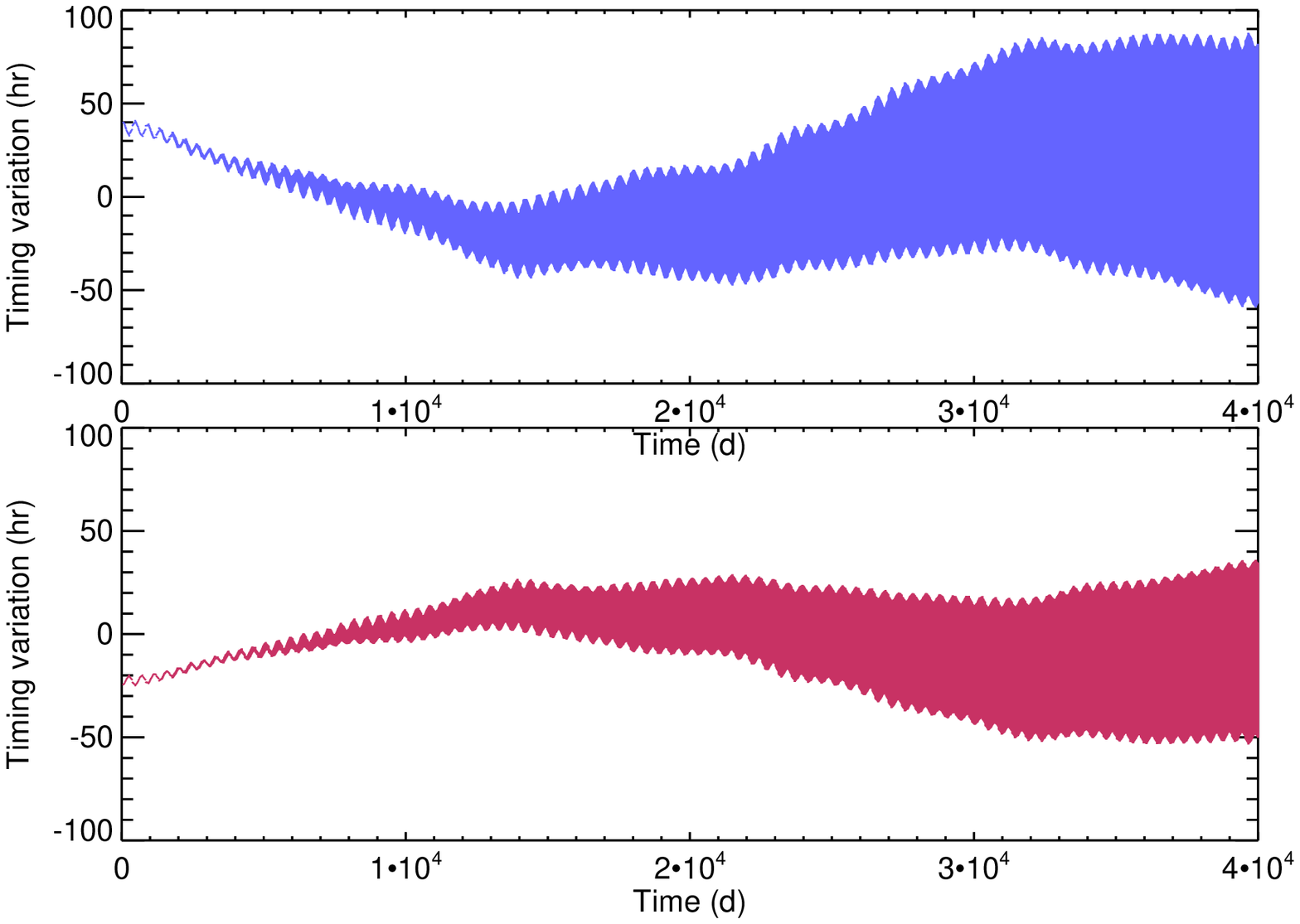}
\caption{Predicted transit timing variations based on the models drawn from the posterior
distribution.  The range indicates the 68\% confidence range predicted for planet b (blue)
and planet c (red).}  \label{ttv_predicted}
\end{figure}

\begin{figure}
\centering
\includegraphics[width=6.5in]{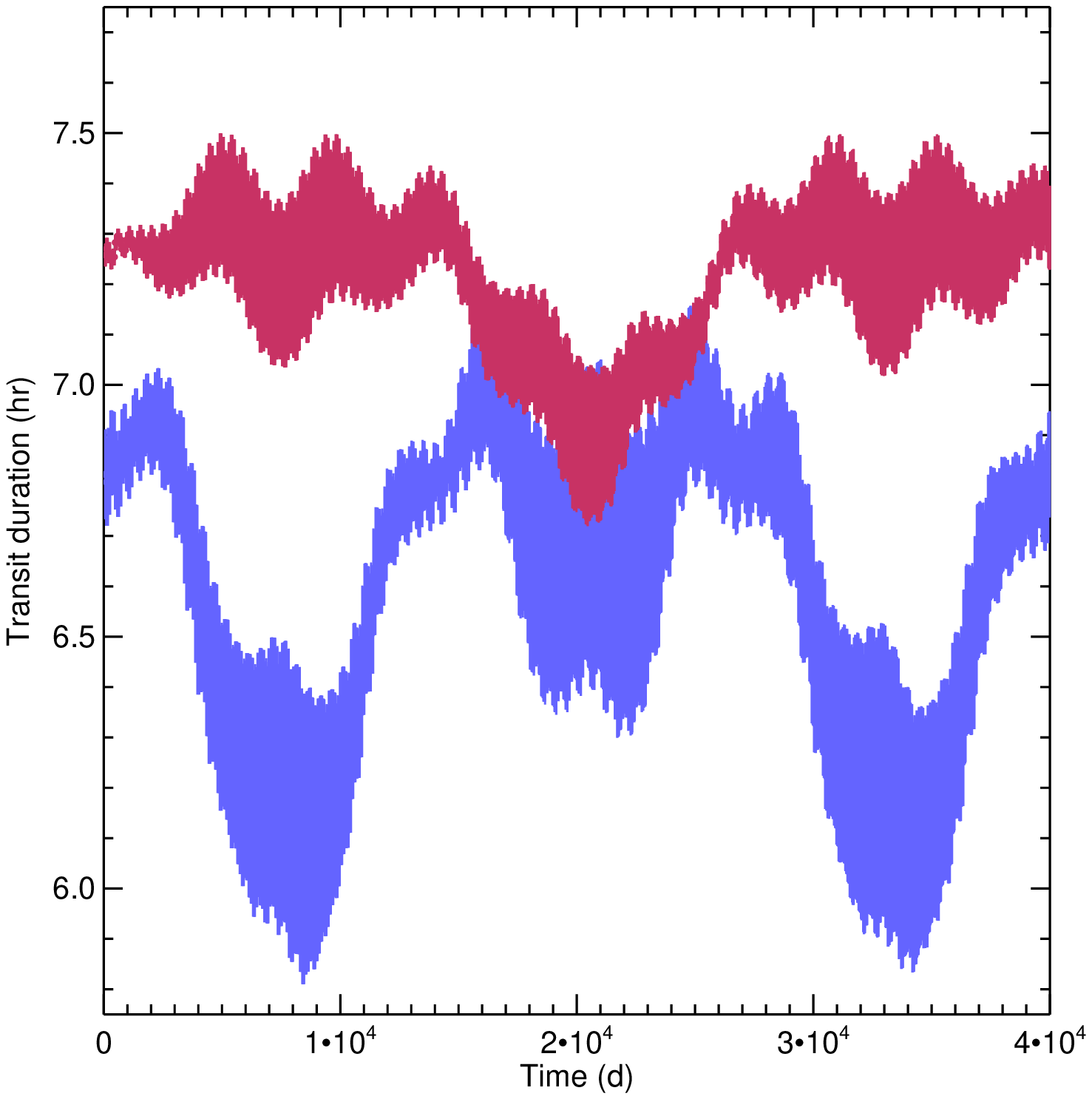}
\caption{Predicted transit duration variations based on the models drawn from the posterior
distribution.  The range indicates the 68\% confidence range predicted for planet b (blue)
and planet c (red).}  \label{tdv_predicted}
\end{figure}

\section{Long-term stability of solutions} \label{sec:stab}

To assess the long term stability of the system, we selected 10,001
 random draws from the posterior distribution of parameters (masses, initial 
 positions, and initial velocities) and integrated each of these for $6.845
 \times 10^5$~years (over 15 million orbits of planet c).  We used a
 symplectic $n$-body map \cite{Wisdom1021991, Wisdom1312006} in
 Jacobi coordinates, with the third-order symplectic corrector
 developed by J. Chambers and reported by Wisdom (2006). In separate
 work we determined that for the mass ratios of Kepler-36b and c (relative to the mass of the star), the
 Chambers corrector provides the optimal balance between precision
 and computational cost.  

The integrations used a time step of 0.125 days.  With the symplectic
corrector,  the relative energy error of the integrations was
typically $10^{-12}$ or smaller,  with a negligible secular trend.
With this time step size, the conjunctions are resolved well. Furthermore, given the degree of conservation of the total energy,
we believe that this time step is small enough to integrate the 
orbits accurately, despite the relatively close encounters that occur. 
To test this, we integrated a small sample of the initial conditions with a 
Bulirsch-Stoer integrator and compared the result
with that from the symplectic $n$-body map. We confirmed that the
output of the two methods agrees for short integrations.

Out of the full set of 10,001 initial conditions, no initial configuration which satisfied the Hill criterion was found to have a close encounter, which we defined as occurring when the planets were within one Hill radius \cite{Hamilton921991} of each other. We selected a random sample of 100 initial conditions and integrated these for $\sim 140$ Million years. Again, there were no close encounters. 

Out of the full set of 10,001 initial conditions, 1,171 satisfied  $(p/a)-(p/a)_{crit} < 10^{-4}$. Of this subset, 313 were formally Hill stable and 858 did not satisfy the Hill criterion [$(p/a)-(p/a)_{crit} < 0$,\cite{Marchal261982}] . In other words, these are the initial conditions that do not, or nearly do not, satisfy the Hill stability criterion. We integrated these for $13.69\times 10^6$ years, again with a time step of 0.125~days and
with a close encounter threshold of $1.0~R_H$.  No initial condition
that satisfied the Hill stability criterion underwent an encounter
during the integrations.  This both further confirms the criterion and supports the
validity of our integrations. 

The majority of the 858 initial conditions that do not satisfy the Hill criterion have close encounters: 616 passed within one Hill sphere of each other during the $13.69$ million years integrations. No initial conditions had a close encounter earlier than $10^6$ days into the integration. Although the Hill stability criterion makes no formal statement about the
stability or instability of orbits that do not satisfy the criterion,
Gladman (1993) \cite{Gladman1061993} found that systems with small eccentricities that do not satisfy the stability criterion
typically undergo close encounters on short time scales. Since the data constrain eccentricities to be relatively small, we might expect that the remaining initial conditions that do not satisfy Hill stability will undergo encounters during longer integrations. There is no pressing need to follow up on these exhaustively, though, as the goal of this stability analysis was not to understand completely the stability of every initial condition. Instead, we have established that $\sim 91 \%$ of the posterior distribution is long lived and does not undergo close encounters. Furthermore, we do not believe that the presence of dynamically unstable orbits in the posterior reflects on the stability of the best fit solution, as the 
Hill unstable portion of the posterior lies on the fringes of the distribution (see Figure \ref{fig:eccomega}).

Although the vast majority of the sample of initial conditions are Hill
stable, this stability criterion alone does not imply that the orbits are
Lagrange stable ({\it i.e.} that their semimajor axes are bounded).
The condition of Lagrange stability can further constrain the posterior, just as the condition of Hill stability does. This will be explored in future work.

\begin{figure}
\centering
\includegraphics[width=5in]{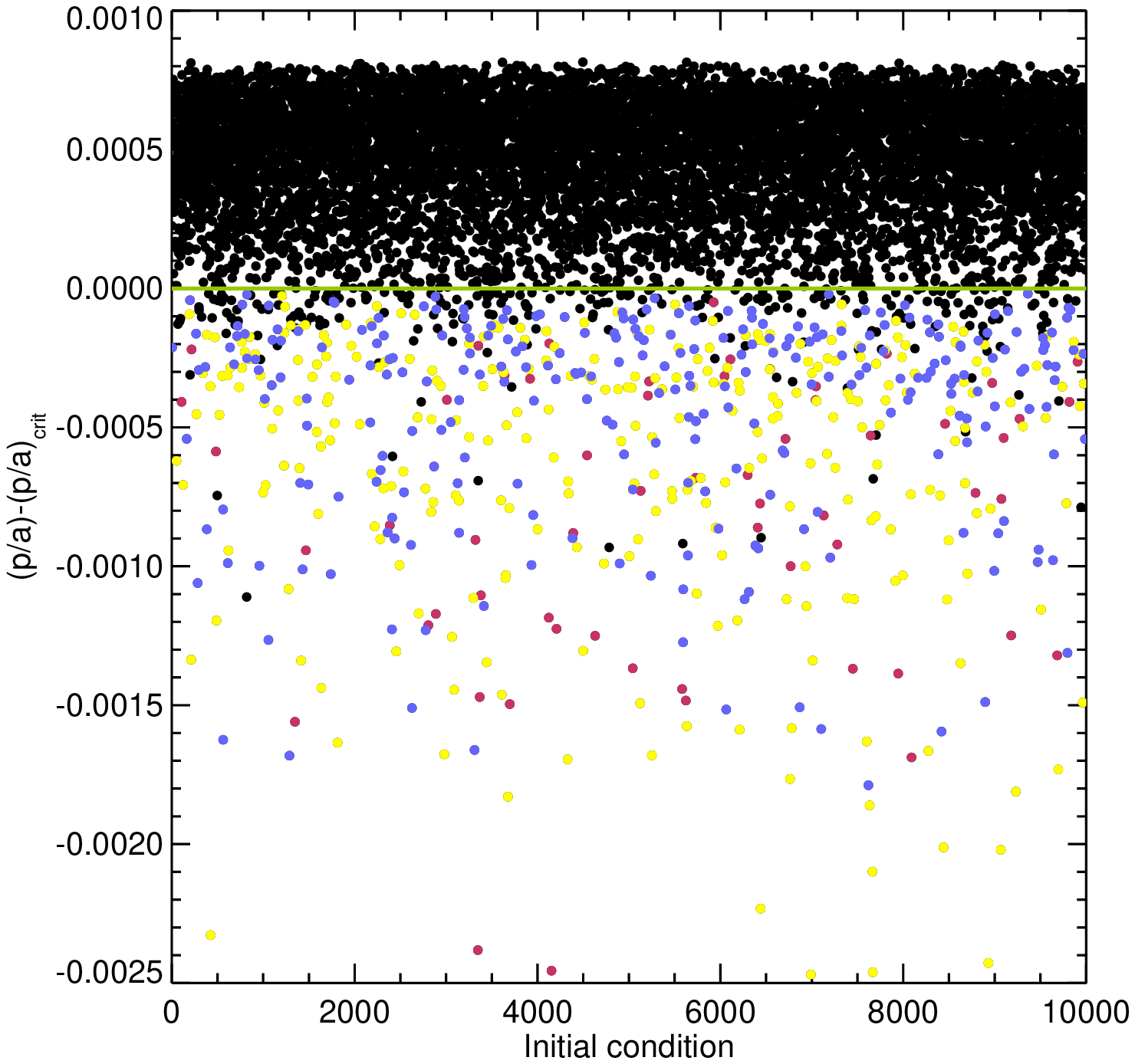}
\caption{Hill and numerical stability diagram for the $10^4$ sets of initial conditions (ICs)
drawn from the posterior distribution of the photo-dynamical
model.  The vertical axis is the dimensionless analytic Hill-stability criterion
due to Marchal \& Bozis (1982) \cite{Marchal261982}:  positive values are Hill stable;  horizontal
axis is the index for the ICs we tested.  The
black dots indicate ICs that were stable after numerical integration
for $2\time 10^{10}$ steps of 0.125 d per step (6.84 Myr).
The colored dots indicate the ICs that became unstable via close approach:
blue became unstable  after $> 3\times 10^8$ time steps, yellow went unstable between
$3\times 10^7-3\times 10^8$ time steps, and red went unstable in $<3\times
10^7$ time steps. }
\end{figure}

 \section{Comparison with other planet systems} \label{sec:comp}

\begin{figure}
\centering
\includegraphics[width=6.5in]{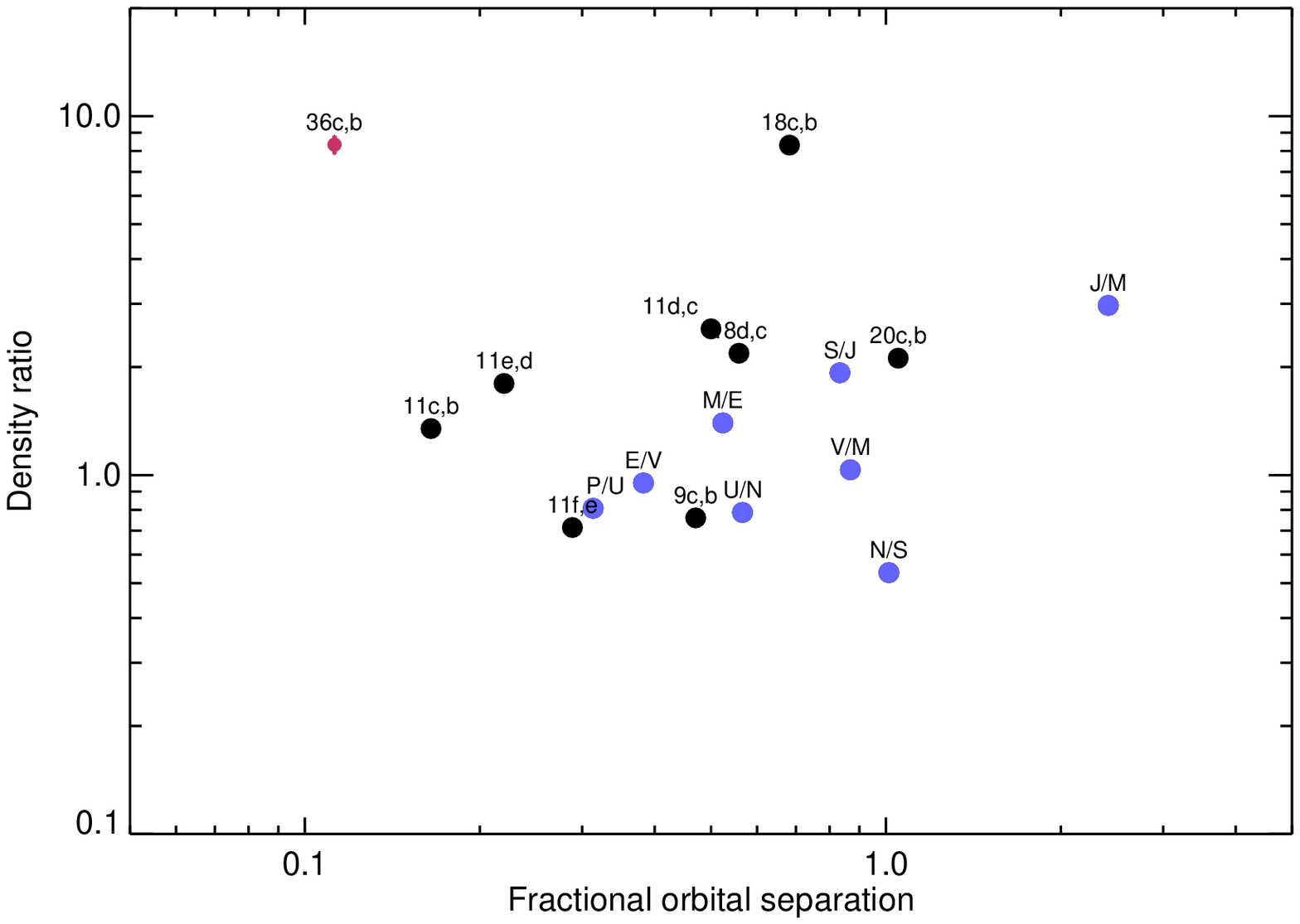}
\caption{Distribution of fractional separation difference ($a_2/a_1-1$) and density ratios
($\rho_2/\rho_1$) for adjacent planets in
our Solar System (blue) and in Kepler systems (black).  Error bars are omitted
since they are smaller than the size of the point for the Solar System, and
are difficult to estimate for the Kepler systems (with the exception of Kepler-36).
The labels use the first letter of the solar system planets, and the Kepler planet number,
(i.e. 11e,b plots the ratio of Kepler-11f to Kepler 11e).  Kepler-36 is plotted
as a red point.}  \label{separation_density}
\end{figure}

Kepler-36 is an outlier in both separation and difference in density:  it has the 
smallest fractional separation of any two adjacent planets with measured masses and 
radii, and it has the largest density contrast, save for Kepler-18b,c (Figure 
\ref{separation_density}). 

\begin{figure}
\centering
\includegraphics[width=6.5in]{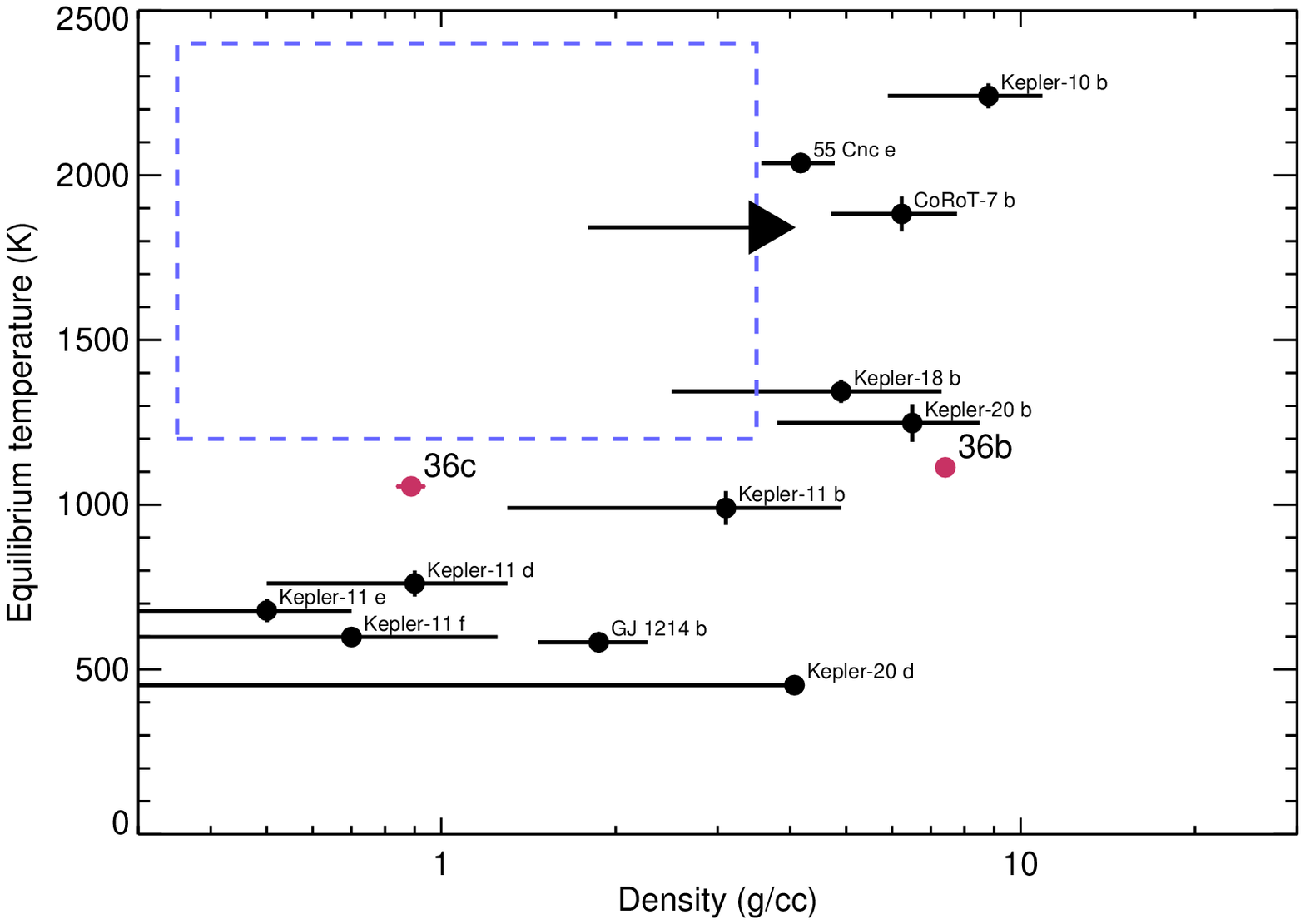}
\caption{Distribution of densities and effective temperatures for super-Earths and
mini-Neptunes ($M_p < 10 M_\oplus$).  We have indicated the lack of hot
mini-Neptunes ($\rho < 3.5$ g/cc, $T < 1200 K$) with a dashed blue box.}  \label{density_temperature}
\end{figure}

The planets Kepler-36 appear to be located at a border between super-Earths
and mini-Neptunes:  Kepler-36b is the coolest super-Earth with a measured
density known to date, while Kepler-36c is the hottest mini-Neptune 
(\ref{density_temperature}).  
In \deletion{both figures}\addition{the figure} we have indicated a suggested division between super-Earths
and mini-Neptunes at a density of 3.5 g/cc;  planets less than this density
likely have a significant H/He envelope.  Any low-density planets (i.e.
mini-Neptunes) at small orbital separations should be easily detected as these 
have a higher transit probability and deeper transit depth.  
It is possible that these planets were formed here, but evaporation
removed the H/He envelopes, causing these planets to
evolve towards super-Earths;  we have indicated this with an arrow in
\deletion{each}\addition{the} figure.  However, it is quite possible that the hot mini-Neptune
region will be filled in with further observations.

\section{Availability of data} \label{sec:avail}

\addition{The {\it Kepler} light curve data used in this analysis can be downloaded from the Mikulski Archives for Space Telescopes at the address below by searching for 11401755 in the ``Kepler ID'' field:} {\tt \begin{verbatim} http://archive.stsci.edu/kepler/data_search/search.php \end{verbatim}}

The spectra used in this analysis have been provided (as ASCII text files) as an attachment to this supplement; 

McDonald co-added spectrum :{\tt \begin{verbatim}1223269s2.txt \end{verbatim}} 

HIRES spectrum: {\tt \begin{verbatim}1223269s3.txt \end{verbatim} }

We have attached the 10,001 initial conditions used in the stability analysis (\S \ref{sec:stab}) (as an ASCII file): {\tt \begin{verbatim}1223269s4.txt \end{verbatim}} 

We have attached a draw of 10,000 model parameters from our full posterior (as an ASCII text file - with FITS encoding for easier input): {\tt \begin{verbatim}1223269s5.txt \end{verbatim}} 

The authors welcome requests for additional information regarding
the material presented in this paper.

\bibliographystyle{Science}

\end{document}